%% file: main.tex
\documentclass[conference]{IEEEtran}
\IEEEoverridecommandlockouts

\newif\ifArxiv
\Arxivtrue

\newif\ifRelease
 \Releasetrue

\usepackage{cite}
\usepackage{amsmath,amssymb,amsfonts}
\usepackage{algorithmic}
\usepackage{graphicx}
\usepackage{textcomp}
\usepackage{xcolor}
\usepackage{url}
\def\BibTeX{{\rm B\kern-.05em{\sc i\kern-.025em b}\kern-.08em
    T\kern-.1667em\lower.7ex\hbox{E}\kern-.125emX}}

\usepackage{tikz}
\usepackage{multirow}
\usepackage[frozencache,cachedir=minted-cache]{minted}
\usepackage{forest}
\usepackage{cleveref}
\usepackage{balance}
\usepackage{booktabs}
\usepackage{xspace}
\usepackage{pgfplots, pgfplotstable}
\pgfplotsset{compat=1.17}   
\usepgfplotslibrary{groupplots,statistics}
\usepackage{mathtools}
\ifRelease
\usepackage[disable]{todonotes}
\else
\usepackage{todonotes}
\fi
\usepackage[scaled=0.8]{FiraMono}
\usepackage{bold-extra}

\definecolor{OliveGreen}{rgb}{0,0.6,0}

\newcommand{\ourTool}{SMASH\xspace}
\newcommand{\ourBm}{MEAMBench\xspace}

\newcommand{\rp}[1]{\todo[color=lime, inline]{rp: #1}}

\newcommand{\co}[1]{\todo[color=orange, inline]{co: #1}}

\newtheorem{theorem}{Theorem}[section]

\newtheorem{example}[theorem]{Example}

\ifRelease

\newcommand{\change}[1]{{#1}\xspace}
\else

\newcommand{\change}[1]{{\color{red}#1}\xspace}
\fi

\newcommand{\nop}[1]{}
\newcommand{\Att}{\mathit{Att}}

\begin{document}

\title{Selective Use of Yannakakis’ Algorithm to Improve Query Performance: Machine Learning to the Rescue}

\author{\IEEEauthorblockN{Daniela Böhm}
\IEEEauthorblockN{Matthias Lanzinger}
\IEEEauthorblockN{Reinhard Pichler}
\IEEEauthorblockN{Alexander Selzer}
\IEEEauthorblockA{
\textit{TU Wien}\\
Vienna, Austria}
\and
\IEEEauthorblockN{Georg Gottlob}
\IEEEauthorblockN{Davide Longo}
\IEEEauthorblockA{
\textit{University of Calabria}\\
Rende, Italy}
\and
\IEEEauthorblockN{Cem Okulmus}
\IEEEauthorblockA{
\textit{Paderborn University }\\
Paderborn, Germany}
}

\maketitle

\begin{abstract}
 Query optimization has played a central role in database research for decades. However,
 more often than not, the proposed optimization techniques lead to a performance improvement
 in some, but not in all, situations. Therefore, we urgently need a methodology for 
 designing a decision procedure that  decides for a given query whether the optimization technique should be applied or not. 
 In this  work, we propose such a methodology with a 
 focus on Yannakakis-style query evaluation as our optimization technique of interest.
 More specifically, we formulate this decision problem as an algorithm selection problem and
 we present a Machine Learning based approach for its solution. Empirical results with 
 a variety of database systems show that 
 our approach indeed leads to a statistically significant \change{and practically notable} performance improvement.
\end{abstract}


\section{Introduction}
\label{sect:Introduction}

 Query optimization has played a central role in database research for decades. 
Some generally accepted techniques 
such as replacing Cartesian product plus selection by a join or projecting out attributes not
needed further up in the query plan as soon as possible are guaranteed to (almost) always lead to 
a performance improvement. However, more often than not, optimization techniques proposed in the literature 
lead to a performance improvement in {\em some}, but not in  {\em all}, situations.
Moreover, it is usually a non-trivial task to delineate the situations where the application 
of a specific optimization is advantageous and where it is not. This also applies to 
optimization techniques which, in theory, should (almost) always outperform the 
conventional query evaluation methods. 

A prominent example of an optimization technique is 
Yannakakis' algorithm~\cite{DBLP:conf/vldb/Yannakakis81}
in case the query is acyclic. 
Several applications and extensions of this algorithm 
in recent time (see, e.g.,%
~\cite{DBLP:conf/sigmod/IdrisUV17,%
DBLP:journals/vldb/IdrisUVVL20,%
DBLP:journals/corr/abs-2301-04003,%
DBLP:journals/corr/abs-2406-17076,%
DBLP:journals/pvldb/BirlerKN24,%
DBLP:journals/corr/abs-2504-03279,%
DBLP:journals/pacmmod/0005023,%
DBLP:conf/sigmod/0001022,%
DBLP:conf/sigmod/Dai0023,%
DBLP:conf/cidr/YangZYK24})
witness the renewed interest in this approach.
The key idea of this algorithm is to first eliminate all dangling tuples (= tuples that will not 
contribute to the final result of the query) via semi-joins and then compute the join with the guarantee that 
all intermediate results thus obtained will be extended to tuples in the final result. Hence, in a sense, 
Yannakakis' algorithm solves the (in general NP-complete) 
join ordering problem that aims at avoiding the 
explosion of intermediate results. In theory, such a strategy of {\em completely avoiding} the generation of 
useless intermediate results should always be superior to conventional techniques that just try to 
compute the joins in an optimal order and thus aim at {\em minimizing} useless intermediate results. 
However, in practice, it turns out that Yannakakis' algorithm leads to a performance improvement in 
{\em some} cases but, by no means, in {\em all} cases.

In this work, we revisit a sub-class of acyclic queries called 0MA (= zero materialization aggregate) 
queries~\cite{DBLP:journals/corr/abs-2303-02723}, 
where the theoretical advantage of Yannakakis-style query evaluation is even more extreme: 
0MA-queries are a restricted class of join queries with an aggregate on top and which can be 
evaluated by carrying out only semi-joins, i.e., completely avoiding the need for computing any joins. 
More precisely, after traversing the join tree $T$ of such a query by semi-joins in bottom-up direction, 
the result can be computed from the relation resulting at the root node of $T$.
(Formal definitions of acyclic queries, 0MA queries, etc.\ are given in 
Section~\ref{sect:Preliminaries}.)
Note that the query given in Figure~\ref{fig:tpch-query}, which is a slightly modified version of a query from 
the STATS benchmark~\cite{DBLP:journals/pvldb/HanWWZYTZCQPQZL21}
(the main modification is that we have replaced \texttt{COUNT}(*) in the \texttt{SELECT} clause by a \texttt{MIN}-expression), falls into this class.
Indeed, if we consider a join tree of this query with the \texttt{comments} relation at the root node, then
we can evaluate this \texttt{MIN}-expression after the bottom-up traversal with semi-joins by only considering the 
resulting relation at the root node.

\begin{figure}[t]
    \centering
    \begin{minted}[escapeinside=||,fontsize=\small]{sql}
SELECT  MIN(|c|.Id) 
FROM    comments AS |c|, posts AS p, votes AS v, users AS u
WHERE   u.Id = p.OwnerUserId AND u.Id = |c|.UserId AND 
   u.Id = v.UserId AND u.Views|$>=$|0 AND p.Score|$>=$|0 AND
   p.Score|$<=$|28 AND p.ViewCount|$>=$|0 AND p.ViewCount|$<=$|6517 
   AND p.AnswerCount|$>=$|0 AND p.AnswerCount|$<=$|5 AND 
   |\colorbox{orange}{p.FavoriteCount$>=$0}| AND p.FavoriteCount|$<=$|8 AND
   |c|.CreationDate|$>=$|'2010-07-27 12:03:40' AND
   p.CreationDate|$>=$|'2010-07-27 11:29:20' AND
   p.CreationDate|$<=$|'2014-09-13 02:50:15' AND
   u.CreationDate|$>=$|'2010-07-27 09:38:05'
    \end{minted}
    \caption{Slightly modified query 121-097 from the STATS benchmark: original runtime on PostgreSQL (3.38s)
    vs.\ 
    Yannakakis-style evaluation (0.11s).
    After changing the filter condition to \colorbox{orange}{\textnormal{\texttt{p.FavoriteCount$>=$8}}}: 
    original runtime on PostgreSQL (0.05s) vs.\ Yannakakis-style evaluation (0.09s).
    }
    \label{fig:tpch-query}
    \vspace{-6mm}
\end{figure}

In theory, it is ``clear'' that such a join-less evaluation must always outperform conventional query 
evaluation techniques that first fully evaluate the underlying join query and only 
then apply the aggregate
as a kind of post-processing. Alas, empirical evaluation 
on queries from several benchmarks shows that, in practice,  this is not necessarily the case.
Actually, for the query in  Figure~\ref{fig:tpch-query}, Yannakakis-style evaluation was significantly faster (by a factor of ca.\ 30) than the original evaluation method of PostgreSQL. However, when we modified one of the conditions in the WHERE clause
from $\geq 0$ to $\geq 8$ (which, together with the $\leq 8$ condition,
yields $= 8$), then suddenly the 
original evaluation method of PostgreSQL is faster. 

In order to reap the fruits of optimizations such as Yanna\-ka\-kis-style query evaluation, 
we urgently need a method 
to design a decision procedure that decides when to apply 
the optimization technique and when to stick to the original evaluation method of the 
database management system (DBMS). We therefore aim at developing a methodology for the design of such a decision procedure, with a focus
on Yannakakis-style query evaluation as our optimization technique of interest.
More specifically, on one hand, we consider general acyclic queries 
(which join all relations after the semi-join-based elimination of dangling tuples)
and, on the other hand, we consider 0MA queries (which only require a bottom-up traversal of the 
join tree with semi-joins). 
\nop{lieber den folgenden Satz unter ``contributions verwenden:
However, 
we believe that the methodology developed here is applicable to a much wider range of query optimization techniques 
that may lead to a performance improvement in some but not in all cases.
}

To cover a diverse set of database technologies, 
we study three quite different DBMSs, namely
PostgreSQL~\cite{DBLP:journals/cacm/StonebrakerK91} (as a ``classical'' row-oriented relational DBMS), 
DuckDB~\cite{DBLP:conf/sigmod/RaasveldtM19} (as
a column-oriented, embedded database), and 
Spark SQL~\cite{DBLP:journals/cacm/ZahariaXWDADMRV16} 
(as a database engine specifically designed for distributed data processing in a cluster).  
This raises the question how to test an optimization strategy for different DBMSs with reasonable effort. 
We therefore adapt previous query rewriting approaches 
on SQL-level (see, e.g.,~\cite{DBLP:journals/corr/abs-2303-02723,%
DBLP:journals/corr/abs-2504-03279,DBLP:conf/amw/GottlobLLOPS23,%
lanzinger2024soft,%
DBLP:journals/pvldb/Zhou0WLSZ23}).
That is, we  take a SQL-query as input and rewrite it to an 
equivalent {\em sequence of} SQL-statements 
that forces (or, at least, guides) 
the DBMS towards Yannakakis-style query evaluation.

We treat the design of a decision procedure between different 
query evaluation techniques as an {\em algorithm selection problem},
that we want to solve by applying Machine Learning (ML) techniques. 
In recent years, many works have used ML techniques to solve database 
problems -- above all cardinality estimation and join order optimization 
(see, e.g., the survey paper~\cite{DBLP:journals/tkde/ZhouCLS22}. 
We will elaborate more on related work in general, and on ML techniques applied to database problems in particular,
in Section \ref{sect:RelatedWork}). 
\change{Our focus, however, is different: we are interested in the 
overall runtime of query evaluation under different evaluation methods.
We ultimately aim to improve end-to-end runtimes over entire workloads, 
rather than local optimizations of some specific subtask of query evaluation.}
%

\nop{***********************
\rp{Reicht/Passt das als Abgrenzung zu anderen ML methoden in DB?   
bzgl. Literatur zu AI in DB habe ich nur das AI4DB survey paper zitiert.}
\co{das ai4DB ist eigentlich kein survey, sondern ein tutorial (4 Seiten). Eine ähnliche gruppe an autoren hat dann später ein survey daraus gemacht, das wird im related works zitiert (21 seiten). Ich hab jetzt mal die Zitate ausgetauscht. Wenn man noch mehr surveys findet, umso besser. }
***********************}

In the first place, we have to fix the specifics of  {\em our} algorithm selection problem. 
This includes the selection of queries and data from benchmarks as well as the identification of relevant 
features that characterize these queries. \change{The input to this ML task is formed by a collection of 
feature vectors (characterizing the selected queries and databases) 
together with runtimes obtained on our target DBMSs  
for the original query evaluation method and for the query optimization method we wish to test. } 

In order to solve this ML task, several decisions have to be made and 
obstacles have to be overcome. For instance, benchmarks aim at testing specific features
of DBMSs and are relatively small
for training and testing ML models. We therefore create a new dataset by 
adapting and combining several benchmarks followed by data augmentation methods 
to achieve the variety and size of the dataset that makes it suitable
for our concrete algorithm selection problem. 
Apart from choosing suitable ML model types together 
with appropriate choices for their hyperparameters,
we have to make the basic decision how to actually define the ML task, 
e.g.: as classification (which evaluation method to choose)
or as regression (to predict the runtime difference between the two evaluation methods, which is then used as basis for the decision between the 
two query evaluation methods). We also have to define the precise criteria
for determining the ``best'' ML model, e.g.: purely quantitative criteria
(such as highest accuracy and precision in case of classification or 
mean squared error in case of regression) or qualitative criteria (which require an inspection of the misclassifications produced by each ML model
and analyzing, for instance, by how much the chosen evaluation method was 
slower than the optimal one). 
%

\smallskip

\noindent
{\em Contributions.} Our main contribution is the 
development of a methodology for designing a decision procedure 
that chooses between different query evaluation techniques.
 We thus present an approach for tackling the design 
    of a decision procedure between different query evaluation techniques 
    as an algorithm selection problem. This approach includes several steps:



\begin{itemize}   
    \item We present an approach for solving this algorithm selection problem
    via ML methods. This includes several substeps such as determining suitable ML model types,
    identifying relevant features, and defining selection criteria for determining the
    ``best'' model. In this work, we focus on optimization techniques based on 
    Yannakakis' algorithm. However, we believe that the 
    methodology of defining and solving the algorithm selection problem
    has the potential for a much wider applicability.

    \item 
    We introduce a new dataset and benchmark -- focusing on hard join queries -- which is extensive and diverse enough for effective model training and testing. This new  benchmark, which we will call
    \textbf{\ourBm} (\textit{\textbf{M}aterialization\textbf{E}xplosion\textbf{A}ugmented\textbf{M}eta\textbf{Bench}mark})
    is constructed by adapting, combining and augmenting queries from several common benchmarks. 
    Beside the algorithm selection problem we are focusing on here, 
    we expect this dataset to be valuable also for other problems, 
    such as query performance prediction.

    \item To prove the practical potential of our approach, we implement it in a system called \ourTool, short for \emph{\textbf{S}upervised \textbf{M}achine-learning for \textbf{A}lgorithm \textbf{S}election  \textbf{H}euristics}. SMASH already supports multiple DBMSs, including PostgreSQL, DuckDB and SparkSQL. It aims at deciding which query optimization method to use for a given query and DBMS.

    \item We report on extensive empirical evaluation that clearly confirms the 
    significance of the improvements achieved by the decision procedure
    compared with the original query evaluation method of the DBMS or 
    always applying the optimization technique.
  
\end{itemize}

\smallskip

\noindent
{\em Structure.}
The paper is organized as follows: After recalling basic definitions and results in 
Section~\ref{sect:Preliminaries}, we give an overview of related work in 
Section~\ref{sect:RelatedWork}.
In Section~\ref{sect:DefiningAlgorithmSelectionProblem}, we formulate
our algorithm selection problem, and we present a methodology for solving this problem in
Section~\ref{sect:SolvingAlgorithmSelectionProblem}. Empirical results are presented in 
Section~\ref{sect:evaluationAlgorithmSelection}. A conclusion and outlook to future work is
given in Section~\ref{sect:conclusion}.

The source code of the implementation, information on running the benchmarks and model training, as well as the data presented here can be found in the following repository: \url{https://github.com/dbai-tuw/yannakakis-rewriting}. 
The \ourBm dataset can be found in the following repository: \url{https://github.com/dbai-tuw/MEAMBench}
\ifArxiv
Some details, which are omitted in the main body of the text, are provided in the 
appendix.
\fi
We also note that the repositories includes various tools, such as Jupyter Notebooks, allowing one to 
rerun experiments themselves, including the training of all tested machine learning models.


\smallskip
\noindent

\section{Preliminaries}
\label{sect:Preliminaries}

\noindent
{\em Conjunctive Queries and beyond.}
The basic form of queries studied here are Conjunctive Queries (CQs),
which correspond to 
select-project-join queries in the Relational Algebra.  
It is convenient to 
consider CQs as Relational Algebra expressions of the form 
$Q = \pi_U (R_1 \bowtie  \dots \bowtie R_n ).$
Here we assume w.l.o.g., that equi-joins have been replaced by 
natural joins via appropriate renaming of attributes. Moreover, we assume that 
selections applying to a single relation have been pushed immediately in front of this relation 
and the $R_i$'s are the result of these selections. 
The projection list $U$ consists of attributes occurring in the $R_i$'s.
\nop{************************
By slight abuse of notation, 
we shall use the same symbol $R_i$ to refer also to the relational schema 
(i.e., the set of attributes) of 
a relation $R_i$. 

\rp{Wir werden sehen, ob wir diesen letzten Satz wirklich brauchen. Er war
jedenfalls im arxiv paper.}
************************}

To go beyond CQs, we will also consider the extension of Relational Algebra by the 
grouping operator $\gamma$ and aggregate expressions. 
In other words, we are interested in queries of the form 
\begin{equation}
\label{eq:basicQueryForm}
       Q  =   \gamma_{g_1, \dots, g_\ell, \; A_1(a_1), \dots, A_m(a_m)} 
       \big( 
      R_1  \bowtie \cdots \bowtie R_n \big)
\end{equation}
where $\gamma_{g_1, \dots, g_\ell, \; A_1(a_1), \dots, A_m(a_m)}$ denotes the
grouping operation for attributes $g_1, \dots, g_\ell$ and aggregate
expressions $A_1(a_1), \dots, A_m(a_m)$.
The grouping attributes $g_1, \dots, g_\ell$ are
attributes occurring in the relations $R_1, \dots, R_n$,
the functions $A_1, \dots, A_m$ are (standard SQL) aggregate functions such as 
MIN, MAX, COUNT, SUM, AVG, etc.,
and 
$a_1, \dots, a_m$ are expressions formed over the attributes 
from $R_1, \dots, R_n$. We have omitted the projection $\pi_U$
in Equation (\ref{eq:basicQueryForm}), since it can be taken care of by the grouping.
A simple query of the form of
Equation (\ref{eq:basicQueryForm}) is given in SQL-syntax in Figure~\ref{fig:tpch-query}.


\smallskip
\noindent
{\em Acyclicity}.
%
An {\em acyclic conjunctive query} (an ACQ, for short) is a 
CQ 
$Q = \pi_U (R_1 \bowtie  \dots \bowtie R_n )$ 
that has a {\em join tree}, i.e., 
a rooted, 
labeled tree $\langle T,r,\lambda\rangle$ with root $r$, 
such that 
(1) $\lambda$ is a bijection that assigns to each node of $T$ one of the relations
    in $\{R_1,   \dots,  R_n\}$ and
(2) $\lambda$  satisfies the so-called {\em connectedness condition\/}, i.e., 
if some attribute $A$ occurs in both relations $\lambda(u_i)$ and $\lambda(u_j)$
for two nodes $u_i$ and $u_j$, then 
$A$ occurs in the  relation $\lambda(u)$ for every node $u$ along the path between $u_i$ and $u_j$.
Deciding if a CQ is acyclic and, in the positive case, constructing a join tree 
can be done very efficiently by the GYO-algorithm
(named after the authors 
of  
\cite{report/toronto/Gra79,DBLP:conf/compsac/YuO79}).
The join query underlying the SQL query in Figure~\ref{fig:tpch-query}
can be easily seen to 
be acyclic. A possible join tree is shown in 
Figure~\ref{fig:tpch-query-jointree}.

\smallskip
\noindent
{\em Yannakakis' algorithm.}
In  \cite{DBLP:conf/vldb/Yannakakis81},
Yannakakis showed that ACQs can be evaluated 
in time $O( (||D|| + ||Q(D)||) \cdot ||Q||)$, i.e., 
linear w.r.t.\ the size of the input and output data and 
w.r.t.\ the size of the query. 
This bound applies to both, set and bag semantics.
Let us ignore grouping, aggregation,  and projection for a while and consider 
an ACQ $Q$ of the form  $R_1 \bowtie \dots \bowtie R_n$ with join tree 
$\langle T,r, \lambda \rangle$. 
Yannakakis' algorithm  (no matter whether 
we consider set or bag semantics) 
consists of a preparatory step followed by 
3 traversals~of~$T$: 

\begin{figure}
    \centering
    \scalebox{1}{
    \begin{forest}
    for tree={align=center}
    [{\texttt{comments}} 
        [{\texttt{users} 
        }            
            [{\texttt{posts} 
            }]
            [{\texttt{votes} 
            }]
        ]
        ]
    \end{forest}
    }
    \caption{Join tree for the query in Fig. \ref{fig:tpch-query}}
    \label{fig:tpch-query-jointree}
    \vspace{-4mm}
\end{figure}
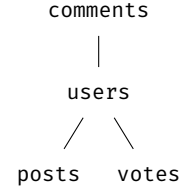

In the {\em preparatory step} 
we associate with each node $u$ in the join tree $T$ the relation 
$\lambda(u)$. If the CQ originally contained selection conditions on attributes of 
relation $\lambda(u)$, then we can now apply this selection. 
The 3 traversals of $T$ consist of 
(1) a bottom-up traversal of semi-joins, (2) a top-down traversal of semi-joins,
and (3) a bottom-up traversal of joins. 
Formally, let $u$ be a node in $T$ with  
child nodes $u_1, \dots, u_k$  of $u$ and let 
relations $R$, $R_{i_1}, \dots, R_{i_k}$ be associated with 
the nodes  $u$, $u_1, \dots, u_k$ at some stage of the computation. 
In the 3 traversals (1), (2), and (3), respectively, 
they are modified as follows:

\smallskip

(1) $R = (((R  \ltimes R_{i_1}) \ltimes R_{i_2}) \dots)  \ltimes  R_{i_k}$,

(2) $R_{i_j} = R_{i_j}  \ltimes R$ for every $j \in \{1, \dots, k \}$, and 

(3) $R = (((R  \bowtie R_{i_1}) \bowtie R_{i_2}) \dots)  \bowtie  R_{i_k}$ 

\smallskip

\noindent
The result 
 of the query is the  final relation associated with 
the root node $r$ of $T$.
Grouping and the evaluation of aggregates can be 
carried out as post-processing {\em after} the evaluation of the join query.
In contrast, projection $\pi_U$ can be integrated {\em into}
this algorithm by projecting out in the second bottom-up traversal all attributes that neither occur
in $U$ nor further up in $T$. Attributes neither occurring in $U$ nor in any join condition 
are projected out as part of the 
preparatory  step.

%

\nop{***********************************
The correctness of Yannakakis' algorithm is seen by 
a closer look at the relations resulting from each 
traversal of $T$. For a node 
$u$ of $T$, let $R$ denote the original relation associated with $u$, i.e., 
$\lambda(u) = R$, and let 
$R_{i_1}, \dots, R_{i_\ell}$ denote the relations
labeling the nodes in the subtree $T_u$ of $T$ rooted at $u$.
Moreover, let $R'$ denote the relation resulting from each traversal of the join tree. 
We again
write (1), (2), (3) to denote the 3 traversals of the join tree.  
Then it holds: 

\smallskip

after (1), we have  $R' = \pi_{\Att(u)} (R_{i_1} \bowtie  \dots \bowtie R_{i_\ell} )$,

after (2), we have  $R' = \pi_{\Att(u)} (R_1 \bowtie  \dots \bowtie R_n )$,

after (3), we have $R' = \pi_{\Att(T_u)} (R_1 \bowtie  \dots \bowtie R_n )$.

\smallskip\noindent
Here, we write $\Att(u)$ and $\Att(T_u)$ to denote the attributes of the 
relation $\lambda(u)$ and the attributes  occurring in any of the relations 
$R_{i_1}, \dots, R_{i_\ell}$
labeling a node in $T_u$, respectively.
***********************************}

\smallskip
\noindent
{\em 0MA queries.}
In \cite{DBLP:journals/corr/abs-2303-02723}, the  class 
of {\em 0MA} (short for {\em ``zero-materialization answerable''}) {\em queries}
was introduced. 
A query of the form given in Equation (\ref{eq:basicQueryForm}) is 0MA if it satisfies the 
following conditions: 

\begin{itemize}
    \item 
{\em Guardedness}, meaning that there exists a relation $R_i$ that 
contains all grouping attributes $g_1, \dots, g_\ell$
and all attributes occurring in the aggregate expressions $A_1(a_1), \dots, A_m(a_m)$. Then $R_i$ is called the {\em guard} of the query.
If several relations satisfy this property, we arbitrarily choose one guard.
\item 
{\em Set-safety}: we call an aggregate function
{\em set-safe}, if it is invariant under duplicate elimination, i.e., 
its value over any set $S$ of tuples remains unchanged
if duplicates are eliminated from $S$. A query satisfies the 
set-safety condition, if all its aggregate functions $A_1 \dots, A_m$
are set-safe.
\end{itemize}

The root of the join tree 
can be arbitrarily chosen. For a 0MA query, we choose the node labeled by the guard
as the root node.
Hence, if all aggregate functions are set-safe (i.e., multiplicities do not matter), 
then we can apply the grouping and 
aggregation $\gamma[g_1, \dots, g_\ell, \; A_1(a_1), \dots, A_m(a_m)]$ to the relation at the root node
right after the first bottom-up traversal.
In SQL, in particular, the \texttt{MIN} and \texttt{MAX} aggregates 
are inherently set-safe.
 Moreover, an aggregate
becomes set-safe when combined with the \texttt{DISTINCT} keyword. For instance,  
\texttt{COUNT}~\texttt{DISTINCT} is 
a set-safe aggregate function.

An example of a 0MA query is given in 
Figure~\ref{fig:tpch-query}:
it is trivially guarded (i.e., there is no grouping and the only aggregate expression is over a single attribute) 
and set-safe (since the only aggregate function 
in this query is \texttt{MIN}).

\section{Related Work}
\label{sect:RelatedWork}

\noindent
{\em Acyclic queries and Yannakakis-style query evaluation.} Yannakakis' algorithm~\cite{DBLP:conf/vldb/Yannakakis81} has recently received renewed attention for the optimization of hard join queries. 
Several works have aimed at bringing its advantages into DBMSs from the outside via SQL query 
rewriting~\cite{DBLP:journals/corr/abs-2303-02723,%
DBLP:conf/amw/GottlobLLOPS23,%
lanzinger2024soft,%
DBLP:journals/corr/abs-2504-03279}, and similar methods such as generating Scala code expressing Yannakakis' algorithm as Spark RDD-operations~\cite{DBLP:conf/sigmod/Dai0023}. 
An important line of research in this area has been concerned with the 
integration of ideas of Yannakakis-style query evaluation into DBMSs while 
avoiding the overhead of {\em several} traversals of the join tree via semi-joins and 
joins~\cite{DBLP:journals/corr/abs-2502-15181,%
DBLP:conf/cidr/YangZYK24,%
DBLP:journals/pvldb/BirlerKN24,%
DBLP:journals/corr/abs-2504-03279}.
\nop{***************************
Even more recently, Yannakakis-like approaches have been proposed, which aim to reduce the overhead of the full Yannakakis' algorithm for enumeration, with its 3 traversals, by instead propagating up additional data and computing the whole query in only one traversal. 
Such approaches have been integrated into Spark SQL~\cite{DBLP:journals/corr/abs-2406-17076}, Umbra~\cite{DBLP:journals/pvldb/BirlerKN24}, DuckDB~\cite{grossadaptive}, and Apache DataFusion~\cite{DBLP:journals/corr/abs-2411-04042}.
***************************}
It should however be noted that, despite all the progress made in integrating and
fine-tuning
Yannakakis' algorithm, we are still left with the fact that this optimization
leads to a performance improvement in {\em some} but {\em not all} cases. 
Indeed, not even for 0MA queries (as one of the simplest forms of acyclic queries),
an improvement in {\em all} cases is guaranteed, as will be confirmed by our empirical 
study presented in  Section \ref{sect:evaluationAlgorithmSelection}.

\nop{*****************************
Further research extends Yannakakis'
algorithm to non-equi-join queries, such as differences of CQs~\cite{DBLP:journals/pacmmod/0005023}, acyclic queries with comparisons 
spanning several 
relations
~\cite{DBLP:conf/sigmod/0001022}, and queries with theta-joins~\cite{DBLP:journals/vldb/IdrisUVVL20}.
*****************************}

%

\smallskip
\noindent
{\em Decompositions.} In order to go beyond acyclic queries, a major area of research seeks to extend Yannakakis-style query answering to "almost-acyclic" queries via various notions of decompositions and their associated width measures, such as hypertree-width, soft hypertree-width, generalized hypertree-width, and fractional hypertree-width~\cite{DBLP:journals/jcss/GottlobLS02,lanzinger2024soft, DBLP:journals/ejc/AdlerGG07,2014grohemarx}. Several implementations~\cite{DBLP:journals/tods/AbergerLTNOR17,DBLP:conf/sigmod/Dai0023,%
DBLP:conf/sigmod/PerelmanR15,DBLP:conf/sigmod/TuR15} combine Yannakakis-style query execution with worst-case optimal joins~\cite{DBLP:journals/jacm/NgoPRR18}. To address the problem of minimal-width decompositions not necessarily being cost-optimal, approaches of integrating statistics about the data into the search for the best decomposition have been proposed and implemented~\cite{lanzinger2024soft, DBLP:conf/pods/ScarcelloGL04}.

\smallskip
\noindent
{\em Query rewriting.} Optimizing queries before they enter the DBMS is a different strategy towards query optimization that has been successfully applied in standard DBMSs~\cite{DBLP:journals/pvldb/ZhouLCF21, DBLP:journals/pvldb/Zhou0WLSZ23}. Although DBMSs already perform optimizations on the execution of the query, it has been shown that rewriting the query itself can still be highly effective.
The WeTune~\cite{DBLP:conf/sigmod/WangZYDHDT0022}  system goes even further, and can be used to automatically discover rewrite rules
but comes with the disadvantage of extremely long runtimes. 

\smallskip
\noindent
{\em Machine learning for databases.} There has been growing interest in the application of machine learning techniques to increase the performance of database systems, as can be seen by a recent survey on this broad area~\cite{DBLP:journals/tkde/ZhouCLS22}. We proceed to give a very brief overview of the general topics 
as to how machine learning has been adapted for database research. For a more detailed account on the rich interaction between machine learning and databases, we refer to~\cite{DBLP:journals/tkde/ZhouCLS22}.
In this survey, the authors categorize the different efforts of using machine learning for core database tasks into several groups. The first group is ``learning-based data configuration''. These are works that aim to utilize machine learning for knob tuning, and view advisor and index advisor tasks~\cite{DBLP:journals/asc/DokerogluBC15,DBLP:conf/sigmod/ZhangLZLXCXWCLR19,DBLP:conf/cloud/ZhuLGBMLSY17,DBLP:conf/sigmod/AkenPGZ17,DBLP:journals/pvldb/LiZLG19,DBLP:journals/pvldb/TanZLCZZQSCZ19,DBLP:conf/cidr/HerodotouLLBDCB11,DBLP:conf/IEEEcloud/NguyenKW18,DBLP:conf/vldb/ChaudhuriN97,DBLP:conf/icde/SchnaitterAMP07}.
Related work that also falls into this category 
is presented in~\cite{DBLP:conf/sigmod/MarcusNMTAK21,DBLP:journals/sigmod/MarcusNMTAK22}.
The next group is ``learning-based data optimization''. These works aim to tackle important, computationally intractable problems such as join-order selection and cardinality estimation of joins~\cite{DBLP:conf/sigir/MizzaroMRU18,DBLP:conf/sigmod/ParkZM20, DBLP:journals/corr/abs-1905-06425, DBLP:journals/corr/abs-1901-08544, DBLP:journals/pvldb/DuttWNKNC19, DBLP:conf/cidr/KipfKRLBK19, DBLP:journals/pvldb/MarcusNMZAKPT19, DBLP:journals/pvldb/MarcusP19,DBLP:conf/sigmod/HeimelKM15, DBLP:conf/vldb/StillgerLMK01, DBLP:journals/tods/TrummerWWMMJAR21, tzoumas2008reinforcement}.
Another group is ``learning-based design for databases''. These works aim more specifically at exploring the use of machine-learning in the construction of various data structures used by modern databases, such as indexes, hashmaps, bloom filters and so on~\cite{ DBLP:conf/sigmod/KraskaBCDP18, DBLP:conf/sigmod/DingMYWDLZCGKLK20,DBLP:conf/sigmod/WuYTSB19, DBLP:conf/sigmod/MaAHMPG18}.
A further group listed in the survey is ``learning-based data monitoring''. As the name suggests, these works aim to use machine learning to create systems that automate the task of running a database and detecting  and reacting to anomalies~\cite{DBLP:journals/pvldb/MaYZWZJHLLQLCP20,DBLP:conf/sigmod/TaftESLASMA18,DBLP:journals/corr/abs-1910-10777,DBLP:journals/pvldb/MarcusP19,DBLP:journals/pvldb/ZhouSLF20}.
Lastly the survey mentions ``learning-based database security''. This category is on how to use 
ML methods to help with critical problems, such as confidentiality, data integrity and availability~\cite{DBLP:conf/kdd/BhaskarLST10,DBLP:conf/icde/ColomboF16a,DBLP:conf/iml/Lodeiro-Santiago17,DBLP:journals/ijcwt/Sheykhkanloo17,DBLP:journals/corr/abs-1901-02868}. 


\smallskip
\noindent
{\em Query performance prediction.} Predicting the performance of a query -- usually the runtime, or sometimes the resource requirements -- is related to the problem of deciding whether to rewrite a query.
Runtime prediction has been performed by constructing cost models based on statistical information of the data \cite{DBLP:journals/is/HeO06}, on SQL queries~\cite{DBLP:conf/icde/WuCZTHN13}, and XML queries~\cite{DBLP:journals/pvldb/ZhouSLF20}. Further approaches use machine learning and deep learning to predict the runtimes of single queries~\cite{DBLP:conf/vldb/ZhangHJLZ05,DBLP:journals/pvldb/MarcusP19,DBLP:conf/sigir/ZhouC07} or concurrent queries (workload performance prediction)~\cite{DBLP:conf/sigmod/DugganCPU11, DBLP:conf/icde/AkdereCRUZ12}.

\begin{figure*}[t] 
  \centering 
    \includegraphics[width=\textwidth]{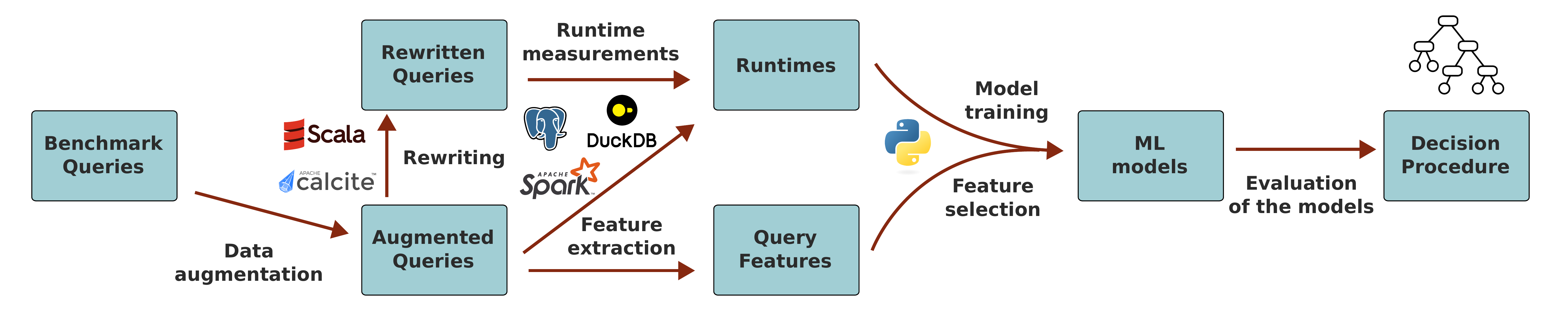} 
  \caption{Methodology workflow.}  
  \label{fig:methodologyWorkflow}  
\end{figure*} 

\section{Formulating the Algorithm Selection Problem}

\label{sect:DefiningAlgorithmSelectionProblem}

Whenever a new, optimized method for query evaluation is presented, we 
need a decision program that decides 
when the new method should be applied. We are thus faced with an algorithm selection 
problem, where we have to decide, for every database instance and query, which query evaluation
method should be applied.  In this section, we describe the steps needed to 
 formulate the precise algorithm selection problem. 
An overview of this workflow is given in Figure~\ref{fig:methodologyWorkflow}. 



It consists of the following steps:
We first have to create a suitable dataset that features the required variety and size for the ML task at hand.
On one hand, we thus have to 
(1) {\em select and adapt common benchmarks} and select those queries to which the optimization technique is applicable. On the other hand, we (2) {\em apply data augmentation} to ensure a suitable size of the dataset.
We then need to (3) {\em select DBMSs} on which we want to test the 
effectiveness of the new optimization techniques.
Implementing these new techniques into a query engine is a non-trivial and expensive task.
Hence, to save effort and to enable testing with various DBMSs,
we make use of the specific nature of Yannakakis-style query evaluation and (4) {\em rewrite the given SQL queries} 
into sequences of SQL commands that ``force'' the 
selected DBMSs into a particular evaluation strategy. 
To prepare for the ML task, we then have to 
 (5)     
{\em do a feature selection}. That is, we want to characterize every query 
in terms of a feature vector. 
We are then ready to (6) {\em run the experiments}, i.e., we execute all queries 
 with and without the  optimization and measure the times of each run. The result will be 
 runtimes for each query (characterized by a particular 
feature vector) both, when the optimization is applied and when it is not applied.
This is then the input to the {\em model training step} and, 
ultimately, to the development of a decision procedure, which will be described 
in Section~\ref{sect:SolvingAlgorithmSelectionProblem}.
%

Below, we describe steps (1) -- (6) in some detail for Yannakakis-style query evaluation of two classes of queries, namely,  {\em enumeration queries} (i.e., queries that ask for the enumeration of the result tuples for an acyclic conjunctive query) and {\em 0MA-queries} (i.e.,  
the subclass of aggregate queries with underlying ACQs~\cite{DBLP:journals/corr/abs-2303-02723}). 
Recall from Section~\ref{sect:Preliminaries}, that 0MA queries allow for a particularly simple evaluation strategy, since they
can be evaluated based on semi-joins only, i.e., without computing any (full) joins. 

\nop{******************************
From a theoretical perspective, one would 
expect the optimization by Yannakakis-style evaluation to be always effective.
This is particularly the case for the 0MA queries, where the performance gain 
achieved by avoiding any join computation should be overwhelming. 
Alas, it turns out that even in this extreme case, the optimization may our may not 
lead to a performance improvement. We thus need to develop decision programs 
for both classes of queries by first formulating this decision problem as an 
{\em algorithm selection problem} and then solving this problem by applying 
ML techniques. Below, we describe the details of the
various steps needed to set up the algorithm selection problem. 
******************************}

\subsection{Benchmark Data} 
\label{sect:benchmarkData}

When constructing our new dataset \ourBm, we pursue two major goals to make the dataset suitable for 
model training and testing: it should be sufficiently big and sufficiently diverse. To address
the diversity aspect,  we collect diverse sets of data and queries from different domains, designed for different purposes, and representing challenging cases of an explosion of intermediate results and materialization.
Thus, as a basis, we have chosen several widely used benchmarks, which contain join queries over several relations: (1) The \textit{JOB (Join Order Benchmark)}~\cite{DBLP:journals/pvldb/LeisGMBK015}, which was introduced to study the join ordering problem, is based on the real-world IMDB dataset and contains realistic join-aggregation queries with many joins and various filter conditions, (2) \textit{STATS/STATS-CEB}~\cite{DBLP:journals/pvldb/HanWWZYTZCQPQZL21} is based on the Stackexchange-dataset, and contains many join queries not following FK/PK-relationships, (3) Four different datasets (namely cit-Patents, wiki-topcats, web-Google and com-DBLP) from \textit{SNAP (Stanford Network Analysis Project)}~\cite{DBLP:journals/tist/LeskovecS16}, \change{a collection of graphs, which we combine with synthetic queries introduced in \cite{DBLP:journals/corr/abs-2406-17076}}, (4) \textit{LSQB (Large-Scale Subgraph Query Benchmark)}~\cite{DBLP:conf/sigmod/MhedhbiLKWS21}, which was designed to test graph databases as well as relational databases, consists of synthetic data and hard queries based on a social network scenario, and (5)
\textit{HETIONET}~\cite{hetionet}. The latter is less known in the database world.
It contains real-world queries on real-world data from a heterogeneous information network of biochemical data, and is part of the new CE benchmark\cite{DBLP:journals/sigmod/ChenHWSS23}, which has, for instance, been recently used 
in~\cite{DBLP:journals/pvldb/BirlerKN24} and \cite{DBLP:journals/corr/abs-2411-04042}.

\nop{*********************
STATS contains a real-world dataset together with a query workload STATS-CEB with  
a number of diverse
multi-table joins, which were specifically designed 
for testing cardinality estimation. 
The SNAP  datasets are commonly used graph datasets. We have decided to use four of them: . The JOB  was designed based on the real-world IMDB dataset to 
test how good query optimizers are. It contains queries with different numbers of joins and filter conditions to test, above all, how well the join ordering works. The LSQB  consists of nine queries based on an underlying graph and allowing to scale up to test query optimizers. The HETIONET dataset is based on a heterogeneous information network of biochemical data with the purpose of connecting multiple different sources into one real-world dataset.
*********************}

We focus on ACQs, which make up the vast majority of the queries in the base benchmarks. 
Most of the queries in the chosen benchmarks are CQs with additional filter conditions 
applied to single tables. These filter conditions can be taken care of by the preparatory step; so they
pose no problem. However, not all of the CQs are acyclic; so we have to eliminate the cyclic ones
from further consideration. 
The number of (acyclic) CQs  of each dataset is given in Table~\ref{tab:augment0MA}.


Note that some of the  queries in the benchmarks are enumeration queries and some already contain some aggregate
(in particular, \texttt{MIN}) and satisfy the 0MA conditions. Of course, also from the enumeration queries, 
we can derive 0MA queries by putting an appropriate aggregate expression (again, in particular, 
with the \texttt{MIN} aggregate function) into the SELECT clause of the query. We do this by 
randomly choosing a table occurring in the query and one column of this table. 
We will see in Section~\ref{sect:DataAugmentation},
that it makes no significant difference which table and attribute we choose for turning a query into 0MA form, 
as we will vary the table and attribute anyway.

\subsection{Data Augmentation}
\label{sect:DataAugmentation}


Our collection of data and queries from different benchmarks results in 219 acyclic queries,
as can be seen in \Cref{tab:augment0MA}. 
Since our goal is to use our new dataset \ourBm for training and testing ML models, this is 
clearly not a sufficient amount. Therefore, we perform data augmentation, as will be detailed next.

For our dataset,  we decided to use the following two steps for data augmentation: 
"filter augmentation" (for all queries) followed by "aggregate-attribute augmentation"
(for 0MA-queries) and ``enumeration augmentation'' (for enumeration queries), respectively. 

With the filter augmentation we want to get duplicates of all queries having filters (i.e., selection conditions on a single table)
and then change some filters in a way that the sizes of the resulting relations vary between these queries. If the query had only one filter we change the specific value it is equal to, greater or smaller of the filter condition. For these cases, we get twice as many queries as before. For all queries having two or more filters we choose two filters, which we change, each at a time. Here we try to replace the filters in a way that once the number of answer tuples gets bigger and once smaller. This gives us triples for each of these queries.

\begin{table}[t] 
\caption{Overview of the 0MA and enumeration queries after augmentation. In total, we get 4677 queries, consisting of 2936 0MA queries and 1741 enum queries. } 
\label{tab:augment0MA} 
\centering
\begin{tabular}{lcccc}
	\toprule
	\bf Dataset &\bf \# ACQs & \bf +filter & \bf +filter\&agg & \bf +filter\&enum\\
	\midrule
	STATS & 146 & 432 & 1876 & 1264\\
	SNAP & \phantom{0}40 & \phantom{0}40 & \phantom{0}244 & \phantom{0}120 \\
	JOB & \phantom{0}15 & \phantom{0}45 & \phantom{0}264 & \phantom{0}135 \\
	LSQB & \phantom{00}2 & \phantom{00}2 & \phantom{00}14 & \phantom{00}6  \\
	HETIONET & \phantom{0}26 & \phantom{0}72 & \phantom{0}538 & \phantom{0}216\\
    \midrule
    \bf Total & 219 & 591 & 2936 & 1741\\
    \bottomrule
\end{tabular}

\end{table}

\begin{example} \label{ex:data_augment1}
Consider the STATS query `005-024', named $q$ here. To illustrate the filter augmentation, we present two possible augmentations on $q$.
One option to augment $q$ is to swap  the filter condition, \mintinline[escapeinside=||,fontsize=\small]{sql}{v.BountyAmount|$>=$|0}  , with a transformed one, such as  \mintinline[escapeinside=||,fontsize=\small]{sql}{v.BountyAmount|$>=$|40}, producing the query $q_{\mathit{aug}1}$. Another option is to swap \mintinline[escapeinside=||,fontsize=\small]{sql}{u.DownVotes=0} with \mintinline[escapeinside=||,fontsize=\small]{sql}{u.DownVotes=10}, producing $q_{\mathit{aug}2}$.

{
\begin{minted}[escapeinside=||,fontsize=\small]{sql}
|{\color{gray}$q$:}|    SELECT MIN(u.Id) 
      FROM   votes AS v, badges AS b, users AS u 
      WHERE  u.Id = v.UserId AND v.UserId = b.UserId 
           AND v.BountyAmount|$>=$|0 AND v.BountyAmount|$<=$|50 
           AND u.DownVotes=0
          
|{\color{gray}$q_{\mathit{aug}1}$:}| SELECT MIN(u.Id) 
      FROM   votes AS v, badges AS b, users AS u 
      WHERE  u.Id = v.UserId AND v.UserId = b.UserId 
           AND |\textbf{v.BountyAmount$>=$40}| AND v.BountyAmount|$<=$|50 
           AND u.DownVotes=0
             
|{\color{gray}$q_{\mathit{aug}2}$:}| SELECT MIN(u.Id) 
      FROM   votes AS v, badges AS b, users AS u 
      WHERE  u.Id = v.UserId AND v.UserId = b.UserId 
           AND v.BountyAmount|$>=$|0 AND v.BountyAmount|$<=$|50 
           AND |\textbf{u.DownVotes=10}|
\end{minted}
}
\end{example}

For 0MA queries, we next apply the "aggregate-attribute augmentation" 
to vary the table from which we take the MIN-attribute. This is done in a way that 
every table occurring in the query appears once in the MIN-expression. 
The column of the chosen table does not really matter, which means we just take the first column of the table. Depending on the number of tables involved in the query, this leads to a different number of new queries per query.

\begin{example} \label{ex:data_augment2}
We give an example for the aggregate-attribute augmentation on 0MA queries. 
As in Example~\ref{ex:data_augment1}, we again focus on the STATS query 005-024. 
For this query, we thus create 3 versions by taking either \texttt{MIN(u.Id)},
\texttt{MIN(v.Id)}, or \texttt{MIN(b.Id)} in the SELECT clause.
This aggregate-attribute augmenation is applied 
to the original queries and to the filter augmented ones alike. 
Hence, the original STATS query 005-024 gives rise to 9 distinct queries after the whole augmentation process.
 \end{example}

\nop{**********************
{\small
\begin{verbatim} 
005-024: SELECT MIN(v.Id) 
         FROM votes as v, badges as b, users as u 
         WHERE u.Id = v.UserId AND v.UserId = b.UserId 
             AND v.BountyAmount>=0 AND v.BountyAmount<=50 
             AND u.DownVotes=0
005-024-augA1: SELECT MIN(b.Id) 
        FROM votes as v, badges as b, users as u 
        WHERE u.Id = v.UserId AND v.UserId = b.UserId 
             AND v.BountyAmount>=0 AND v.BountyAmount<=50 
             AND u.DownVotes=0
005-024-augA2: SELECT MIN(u.Id) 
        FROM votes as v, badges as b, users as u 
        WHERE u.Id = v.UserId AND v.UserId = b.UserId 
             AND v.BountyAmount>=0 AND v.BountyAmount<=50 
             AND u.DownVotes=0
\end{verbatim}
}
**********************}



In Table~\ref{tab:augment0MA}, we summarize the numbers of 0MA queries that 
we get after each step of the augmentation. The SNAP and LSQB queries do not have filter conditions, which means there is no 
filter augmentation for them. 

We also take the enumeration queries, for which filter augmentation has already been done,
and apply an enumeration augmentation step. To this end, we randomly choose two of the 
attributes used in join conditions and write them into the SELECT clause of the query. This is done three times for each filter augmented query if at least three different join attributes exist in the query. 
On the other hand, a query with only one join gives rise to only a single enumeration query (with the 
join attributes in the SELECT clause) in our dataset.

In summary, after applying the data augmentation step to the 0MA and enumeration queries,
we have 4677 queries in total.
\change{
\ifArxiv In \Cref{app:dataAugmentation},
\fi
we give further examples on the data augmentation, namely on the enumeration augmentation. 
}

\subsection{Selection of DBMSs} 
\label{sect:SelectionDBMSs}

We want to check the effectivity of the optimization via Yannakakis-style query evaluation
on
a wide range of database technologies. We have therefore chosen three significantly different DBMSs,
namely 
(1) PostgreSQL 13.4~\cite{DBLP:journals/cacm/StonebrakerK91} as a ``classical'' row-oriented relational DBMS, 
(2) DuckDB 0.4~\cite{DBLP:conf/sigmod/RaasveldtM19} as
a column-oriented, embedded database, and 
(3) Spark SQL~3.3~\cite{DBLP:journals/cacm/ZahariaXWDADMRV16} 
as a database engine specifically designed for distributed data processing in a cluster.  
These DMBSs represent a broad spectrum of architectures and characteristics and  they, therefore, give a good overview of the range of existing DBMSs.

\subsection{Query Rewriting}
\label{sect:QueryRewriting}

To recall our motivation, we aim to find methods that determine the effectiveness of various optimization methods -- with a focus in this work on Yannakakis-style query evaluation  -- on various DBMSs before potentially having to commit to the significant effort of modifying existing query engines. 
Hence, we make use of ideas from recent works~\cite{DBLP:journals/corr/abs-2303-02723,DBLP:conf/amw/GottlobLLOPS23,lanzinger2024soft,DBLP:conf/sigmod/Dai0023} that present query rewritings, where a single SQL query is rewritten into an equivalent series of queries, to guide DBMSs to utilize a given optimization method. 
\change{
\ifArxiv
In \Cref{app:query_rewriting_example},
\fi
we illustrate the query rewriting approach that we use in this paper, using an example query from our benchmark.

}

\subsection{Feature Selection} 
\label{sect:FeatureSelection}

We choose different kinds of features that we derive from the structure of the 
query itself, from the join tree constructed in the process of rewriting the query, and 
from statistics determined by PostgreSQL or DuckDB over the database. The latter kind of features
is extracted from the query optimizer's estimates, and obtained via the EXPLAIN command. 
Note that Spark SQL does not provide an EXPLAIN command. However, we will explain below how to circumvent this shortcoming.
Another challenge are features that are based on a set, of variable length, containing numeric values. In order to reduce such a set of values into a fixed-length list of values, we calculate, for each set, several statistics: min, the 0.25-quantile (referred to as q25), median, 
the 0.75-quantile (referred to as q75), max, and mean.  In the list of features below, we use $^*$ (e.g. \textbf{B7*}) to mark which features consist of variable-length sets, and hence will get reduced to the mentioned collection of 6 values.

\smallskip

\noindent
{\em Features derived from the query.} 
The following features are easily obtained by inspecting the query itself: \\ 
\noindent
\textbf{Feature B1}: \textit{is 0MA?} indicates (1 or 0) if the query is 0MA, \\
\textbf{Feature B2}: \textit{number of relations}, \\
\textbf{Feature B3}:
\textit{number of conditions}, which refers to the number of (in)equality conditions in the WHERE clause of the query, \\
\textbf{Feature B4}: 
\textit{number of filters}, which more specifically only counts the (in)equality conditions occurring in the query, and\\
\textbf{Feature B5}: \textit{number of joins}.

\smallskip

\noindent
{\em Features based on the join tree.} The following features are inspired by the work in~\cite{DBLP:journals/jair/AbseherMW17}
on tree decompositions:  \\
\textbf{Feature B6}:
\textit{depth}, which is the maximal distance between the root of the used join tree and a leaf node, \\
\textbf{Feature B7*}: \textit{container counts}, which is a set of numbers, indicating for each variable in the query the number of nodes in the join tree it occurs in. This measure indicates how many relations are joined on the same variable, and \\
\textbf{Feature B8*}:
\textit{branching degrees}, which is a set of numbers, indicating for each node the number of children it has.

\medskip

These eight features (B1)-(B8*) are the shared {\em "basic features"}.
In addition, we can use statistical information from the database and the estimates for the query evaluation, though the exact features that are exposed differs between DBMSs. In case of PostgreSQL and DuckDB, we
have the EXPLAIN command at our disposal to obtain relevant further information. 
For PostgreSQL, we thus select the following additional features, which we 
refer to as {\em "PSQL features"}:\\
\textbf{Feature P1}: \textit{estimated total cost} (of the query), \\ 
\textbf{Feature P2*}: 
\textit{estimated single table rows}, which stands for the estimated number of rows for each 
table involved in the query {\em after} the application of the filter conditions, and \\ 
\textbf{Feature P3*}:
\textit{estimated join rows}, the estimated  number of rows of each join {\em before} the application of the filter conditions. 
\noindent

\medskip

\noindent
The EXPLAIN command for DuckDB behaves differently. It allows us to 
derive a single {\em "DDB feature"}: \\
\textbf{Feature D1*}:
\textit{estimated cardinalities}, the estimated number of rows after each node in the logical plan, such as  filters and joins.



As SparkSQL does not perform cost-based optimization, it also does not have any statistical information of the data, and cannot estimate cardinalities or costs of a plan.  However, since SparkSQL also does not provide a persistent storage layer, tables are commonly imported from another database via JDBC. This implies that, in practice, the statistical features can easily be extracted from this database and used for the decision whether to rewrite the query in SparkSQL. \change{For the experiments presented in \Cref{sect:evaluationAlgorithmSelection}, we extracted these features from PostgreSQL, hence SparkSQL will have the same feature set as PostgreSQL.}

\ifArxiv
In \Cref{app:FeatureSelection},
\fi
we illustrate the basic features as well as the additional 
features for PostgreSQL and DuckDB 
by looking at two queries from 
the HETIONET benchmark.

\subsection{Running the Queries}

\nop{******************
\begin{figure}[t] 
  \centering 
    \includegraphics[width=0.4\textwidth]{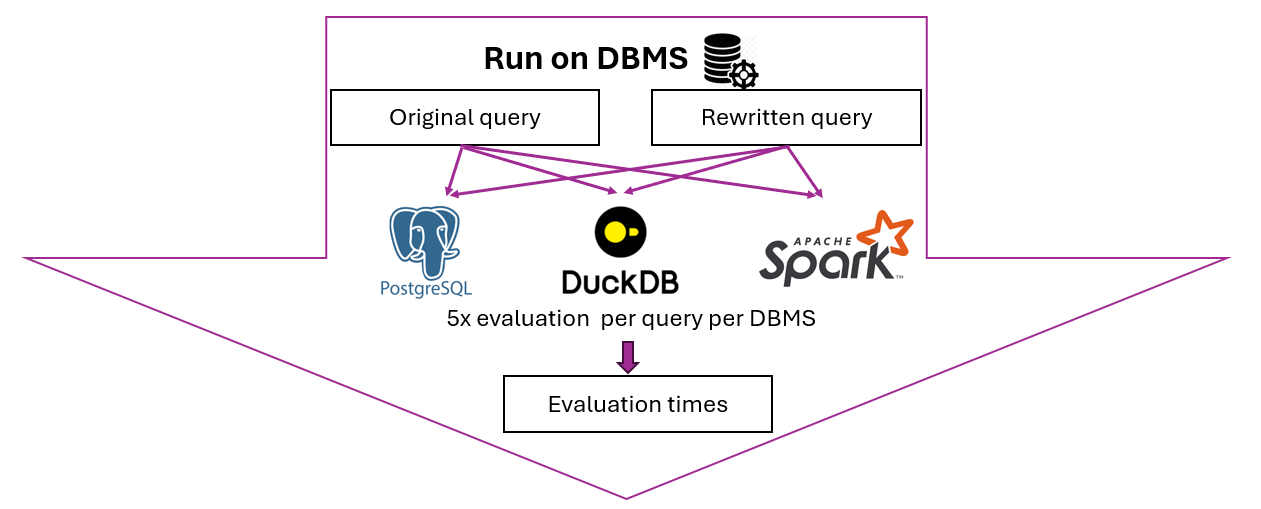} 
  \caption{Workflow of the query evaluation.}  
  \label{implement_dbms}  
\end{figure} 

In Figure~\ref{implement_dbms} we illustrate this part of the workflow. 
******************}
The whole evaluation is performed on a server with 128GB RAM, a 16-core AMD EPYC-Milan CPU, and a 100GB SSD disk, running Ubuntu 22.04.2 LTS. 
After a warm-up run, the original query, as well as the rewritten version, is evaluated five times, and then we take the mean of those five runtimes. 
In total we get 6 data points for each of the 4677 queries: each query is run against 3 DBMSs, where the  query evaluation happens  once with and once without optimization. 
Aggregated information on these 
runtimes is provided in Section~\ref{sect:evaluationAlgorithmSelection}. 

\change{ In addition to the six runtimes, we also add a ``feature vector'', consisting of the features described in \Cref{sect:FeatureSelection}. These provide the 
input to the training of ML models to be described next.
}

\section{Solving  the Algorithm Selection Problem}
\label{sect:SolvingAlgorithmSelectionProblem}

Several decisions have to be made to solve the 
algorithm selection problem that results from the steps 
described in the previous section. In particular, 
we have to (1) {\em formulate the concrete ML task} and 
then (2) {\em select ML model types} together with 
hyperparameters appropriate to our context. Before we can start training and testing the 
models, we have to (3) {\em split the data} \change{(in our case the SQL queries) }
into training/validation/test data. Finally, we have to  (4)
{\em define  selection criteria} for determining the ``best'' model, which 
will then be used as basis of our decision program between the 
original query evaluation method of each system and a Yannakakis-style evaluation.

\subsection{Formulating the ML Task}
\label{sect:FormulatingML:Task}

Our ultimate goal is the development of a decision program between two 
evaluation methods for acyclic queries. Hence, 
we are clearly dealing with a {\em classification problem} with {\em 2 possible outcomes}. 
In the sequel, we will refer to these two possible outcomes as 0 vs.\ 1 to denote
the original evaluation method of the DBMS vs.\ a Yannakakis-style evaluation (enforced
by our query rewriting). 

On the other hand, it also makes sense to consider a {\em regression problem} first and, depending 
on the predicted value, classify a query as 0 (if the predicted value suggests faster 
evaluation by the original method of the DBMS \change{within a certain threshold}) or 1 (otherwise). 
As target of the regression problem we would like to choose the difference 
$t_{\mathrm{rewritten}} - t_{\mathrm{original}}$, where we write  
$t_{\mathrm{rewritten}}$ and $t_{\mathrm{original}}$ to denote the time 
needed by Yannakakis-style evaluation and by the original evaluation method of the DBMS, respectively. 
\change{However, as will become clear in our presentation of the experimental results in Section~\ref{sect:evaluationAlgorithmSelection},
the actual runtime values are very skewed, in the sense that their distribution shows high variance. Hence, the difference we focus on is also  highly skewed. To get more reliable results, we therefore perform a (variant of) log-transformation as 
described next:
Since we may have negative values, we cannot apply the
logarithm directly.} Instead, we multiply the log of the absolute values 
with the sign they had before. Additionally, since we have a lot of values close to zero (which leads to very small log values)
we add 1 to the absolute values before applying the log, which is a common method in such situations. 
The transformation therefore results in the following formula: $x_{new} = sgn(x) * log(|x| + 1)$. \change{ In \Cref{fig:difference_comparison}, we can see this difference function under log transformation over the data from our experiments shown in \Cref{sect:evaluationAlgorithmSelection}.}

\subsection{Selecting the ML Model Types}
\label{sect:SelectingML:Types}

We have chosen 7 Machine Learning model types for our algorithm selection problem, namely
k-nearest neighbors (k-NN), decision tree, random forest, support vector machine (SVM), and 3 forms of 
neural networks (NNs): multi-layer perceptron (MLP), hypergraph neural network (HGNN)
and a combination of the two. 
MLP is the ``classical'' deep neural network type. Hypergraph neural networks, introduced in~\cite{DBLP:conf/aaai/FengYZJG19}, are less known. With their idea of representing the hypergraph structure in a vector space, 
HGNNs seem well suited to capture structural aspects of conjunctive queries. Just like MLPs, also the
HGNNs 
produce an output vector. In our combination of the two model types, 
we provide yet another
neural network, which takes as input the two output vectors produced by the MLP and the HGNN
and combines them to a joint result using additional layers.

A major task after choosing these ML model types is to fix the hyperparameters. An overview of some basic hyperparameters is 
shown in 
Table~\ref{tab:model_hyperparameters}.
Of course, in particular for the 3 types of neural networks, 
many more hyperparameters have to be fixed. 
A detailed
list of all hyperparameter values, in particular for the 3 types of neural networks used here, 
is provided 
\ifArxiv
in \Cref{app:SolvingAlgorithmSelectionProblem}.
\fi

\subsection{Labeling and Splitting the Data}
\label{sect:SplittingData}

After running the 4677 queries mentioned in Section \ref{sect:DataAugmentation} on the 
3 selected DBMSs according to Section \ref{sect:SelectionDBMSs}, we have to prepare the 
input data for training the ML models of the 7 types mentioned in Section \ref{sect:SelectingML:Types}.
Recall from Section \ref{sect:FeatureSelection} that each query is characterized by a 
feature vector specific to each of the 3 DBMSs. For our supervised learning tasks 
(classification and regression), 
we have to label each feature vector for each of the 3 DBMSs. As explained in 
Section~\ref{sect:SelectingML:Types}, we want to train our models both, for classification and 
for regression. Hence, on the one hand, each feature vector gets labeled 0 or 1 
(meaning that the original evaluation of the DBMS or the Yannakakis-style evaluation is faster; 
in case of a tie, we assign the label 0) for the classification
task. On the other hand, each feature vector is labeled with the 
difference of the runtime of the original evaluation minus
the runtime of the Yannakakis-style evaluation for the regression task. 

The labeled data can then be split into training data, validation data,  and test data. 
In principle, we choose a quite common ratio between these three sets by letting the training set contain 80\% of the data and 
the other two contain 10\% each. 
However, to get more accurate results, 
we have decided to do 10-fold cross validation. That is, we split the 90\% of the data 
that were chosen for training and validation in 10 different ways in a ration 80:10 into training:validation data
and, thus,  repeat the training-validation step 10 times.

\subsection{Selection Criteria for the ``Best'' Model}
\label{sect:Selection Criteria}

In order to ultimately choose the ``best'' model for our decision program between the 
original evaluation method of each DBMS and the Yannakakis-style evaluation, we compare, for every feature vector, the predicted classification with the actual labeling. 
We refer to classification 1 as ``positive'' and classification
0 as ``negative''. This leads to 4 possible outcomes of the comparison between predicted and actual value, namely TP (true positive) and TN (true negative) for correct classification and 
FP (false positive) and FN (false negative) for misclassification. They give rise to the 
3 most common metrics: accuracy (shortened to ``Acc''), which is the proportion of correct classifications, precision (shortened to ``Prec''), which is defined by TP / (TP + FP), and recall (shortened to ``Rec''), defined as TP / (TP + FN). 


Of course, the natural goal when selecting a particular model is to maximize the 
accuracy. However, in our context, 
we consider the precision equally important. That is, we find it particularly important 
to minimize false positives, i.e.,  in case of doubt, it is better to stick to 
the original evaluation method of the DBMS rather than wrongly choosing an alternative 
 method.

For regression, we aim at minimizing the mean squared error (MSE). 
But ultimately, we also map the (predicted and actual) difference between the 
runtime of the original minus Yannakakis-style evaluation to a 0 or 1 classification. 
Hence, we can again measure the quality of a model in terms of accuracy, precision, and recall. 

Apart from the purely {\em quantitative} assessment of a model in terms of accuracy, precision, 
and recall, we also carry out a {\em  qualitative} analysis. That is, for each of the misclassified
cases, we want to investigate by how much the chosen evaluation method is 
slower than the optimal method. And here, we are again particularly interested in the
false positive cases. Apart from aiming at high accuracy and precision, we also want to 
make sure for the false positive classifications, that the difference in the 
runtimes between the two evaluation methods is rather small.

\begin{table}[t]
    \centering
    \caption{Performance of Machine Learning Classifiers on the PostgreSQL runtimes. We show accuracy, precision and recall for binary classifiers that predict whether rewriting to Yannakakis style evaluation leads to performance gain.}
    \label{tab:ml.performance_class}
    \setlength{\tabcolsep}{2pt}
    \begin{tabular}{lccc|ccc}
        \toprule
        \textbf{Algorithm} & \multicolumn{3}{c|}{\textbf{0MA Queries}} & \multicolumn{3}{c}{\textbf{Acyc. Queries}} \\
        \cmidrule(lr){2-4} \cmidrule(lr){5-7} & Acc. (\%)$\uparrow$  & Prec. $\uparrow$ & Rec. $\uparrow$ & Acc. (\%) $\uparrow$ & Prec. $\uparrow$ & Rec. $\uparrow$ \\
        \midrule
        Decision Tree & \bf 0.94 & \bf 0.92 & \bf 0.97     & \bf 0.95 & \bf 0.95 & 0.92 \\
        Random forest & \bf 0.94 & \bf 0.92 & \bf 0.97         & \bf 0.95 & 0.94 & \textbf{0.93} \\
        $k$-NN & 0.91 & 0.91 & 0.90                    & 0.91 & 0.88 & 0.91 \\
        SVM & 0.85 & 0.85 & 0.84                       & 0.84 & 0.82 & 0.77 \\
        MLP & 0.87 & 0.89 & 0.86                       & 0.85 & 0.84 & 0.77 \\
        HGNN & 0.83 & 0.84 & 0.85                      & 0.79 & 0.70 & 0.75 \\
        HGNN+MLP  & 0.82 & 0.78 & 0.93                 & 0.81 & 0.77 & 0.72 \\
        \bottomrule
    \end{tabular}
\end{table}

\section{Experimental Results}
\label{sect:evaluationAlgorithmSelection}

In this section, we present experimental results obtained by putting the algorithm selection method described in 
Sections~\ref{sect:DefiningAlgorithmSelectionProblem} and~\ref{sect:SolvingAlgorithmSelectionProblem} to work. We thus first evaluate 
in Section~\ref{sec:mlexp}
the performance of various machine learning models on the raw dataset of query runtimes obtained for the selected and augmented benchmarks on the chosen DBMSs.
Afterwards, in Section~\ref{subsec:effects_perf},
we  evaluate the performance gains by combining the best algorithm selection model with our rewriting method for evaluating acyclic queries.  
In particular, we perform experiments to answer the following key questions:

\begin{description}
    \item[Q1] How well can machine learning methods  predict whether Yannakakis-style query evaluation
    is preferable over standard query execution?
    \item[Q2] Can we use these machine learning models to gain insights about the circumstances in which Yannakakis-style query evaluation is preferable over standard query execution?
    \item[Q3] How well does good algorithm selection performance translate to query evaluation times on 
    different DBMSs?
    \item[Q4] To what extent can we optimize for precision while maintaining end-to-end runtime and accuracy?
\end{description}


In the remainder of this section, we present our experimental results and 
a discussion as to how they answer our key questions. Further details 
\ifArxiv
are provided \change{in \Cref{app:AdditionalExp}.}
\fi

\begin{table}[t]
    \caption{Chosen hyperparameters.}
    \label{tab:model_hyperparameters}
    \centering
    \begin{tabular}{ll}
        \toprule
        \bf Model & \bf Hyperparameters \\ 
        \midrule
        Random Forest & \#estimators = 100 \\
        kNN & k = 5 \\
        SVM & kernel = linear \\
        MLP & layers = 30-60-40-2 \\
        HGNN & layers = 1-16-32-2 \\
        \bottomrule
    \end{tabular}
\end{table}

\begin{table}[t]
    \centering
    \caption{Performance of Regression Models. We show MSE and MAE for regression models predicting the difference between the runtime of the original and the rewritten query. Additionally, we present the accuracy and precision after converting the regression model to a classification model by setting a threshold at a predicted time difference of $0$ seconds.}
    \label{tab:ml.performance_reg}
    \begin{tabular}{lcc|ccc}
        \toprule
        \textbf{Algorithm} & \multicolumn{5}{c}{\textbf{Acyc. Queries}} \\
        \cmidrule(lr){2-6}
        & MSE  & MAE  & Acc. & Prec. & Rec.  \\
        \midrule
        Decision Tree   &  \bf 0.02 & \bf 0.06 & \bf 0.96 & \bf 0.96 & \bf 0.94 \\
        Random forest   & 0.03 & 0.08 & 0.95 & 0.94 & 0.93 \\
        $k$-NN          & 0.17 & 0.18 & 0.91 & 0.88 & 0.91 \\
        SVM             & 1.02 & 0.61 & 0.79 & 0.76 & 0.72 \\
        MLP             & 0.39 & 0.32 & 0.81  & 0.72 & 0.77 \\
        HGNN            & 0.74  & 0.48  & 0.71  & 0.57 & 0.85 \\
        \bottomrule
    \end{tabular}
\end{table}



\subsection{Model Training}
\label{sec:mlexp}

In a first step, we will compare the performance of various learned models in terms of accuracy, precision, and recall. 
\change{Table~\ref{tab:ml.performance_class} compares the performance of the best classification models (with the hyperparameters given in Table~\ref{tab:model_hyperparameters}
on the 0MA queries only, 
as well as on all queries (i.e., 0MA and enumeration), on the runtime data from PostgreSQL. Decision trees and random forests, with roughly the same performance, achieve the best accuracy, precision and recall out of all classifiers. kNN achieves the next best performance, although significantly less at 91\% accuracy compared to the 95\% accuracy of the decision tree-based models. The random forest classifier has a similar performance to the simple decision trees. Hence, its disadvantage of additional complexity is not 
outweighed by a significant performance improvement. Therefore, the simple decision tree classifier is the clear choice out of all compared.
}

\begin{figure}
    \centering

\begin{tikzpicture}
\begin{axis}[
    ymode=log,
    ybar,
    ymin=0,  
    height=4cm,
    width=\columnwidth,
    xlabel style={align=center},
    xlabel=
    $\mathit{diff} \coloneqq \mathit{rewritten} - \mathit{original\_runtime}$ \\    
    $\mathit{sgn}(\mathit{diff}) \log(|\mathit{diff}|+1)$
]
\addplot +[
    hist={
        bins=30
    }   
] table [y index=1, col sep=comma] {data/y1_diff_log.csv};
\end{axis}
\end{tikzpicture}
    \caption{Distribution of the regression response, understood as the time difference between rewritten runtime and original runtime for PostgreSQL, under the given log transformation.  Above we show a histogram of this distribution, using log scaling to allow for more visual clarity. 
    }
    \label{fig:difference_comparison}
\end{figure}
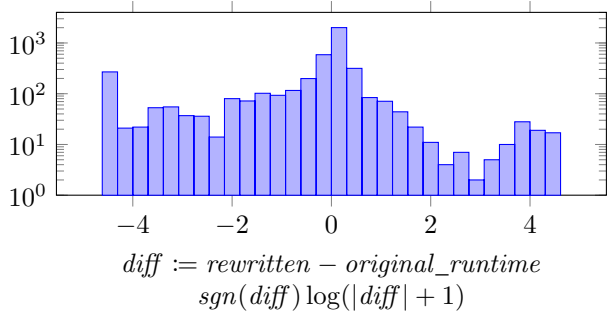

Next, we compare the performance of the regression variants of these models, presented in Table \ref{tab:ml.performance_reg} and
(applying the transformation described in Section \ref{sect:FormulatingML:Task}) 
in Figure~\ref{fig:difference_comparison} 
on all queries, again on the PostgreSQL runtimes.  We can see that the regression performance in terms of MSE and MAE corresponds closely to the classification performance of the classification models, with the decision tree and random forest models again at the top. However, the gap to the next best model is even larger, corresponding to a 5-8x increase in MSE and a 2-3x increase in MAE, making the decision tree-based models again preferable in this situation. 
From these results, it can be observed that we are able to predict the runtime difference (i.e., runtime of the
rewritten query minus runtime of the original query) quite accurately. This predicted runtime
difference naturally lends itself to classification by choosing the query rewriting if and only if
this difference is below 0. This classification derived from the decision tree regression model 
leads to a similar (actually slightly better by 0.5\%) performance than the decision tree classification.
Moreover, as will be discussed next, classification based on the predicted runtime difference provides
additional flexibility for fine-tuning the trade-off between precision and recall. 
Hence, we have chosen classification based on the 
decision tree regression model as basis for our decision procedure.

\begin{figure}[t]
    \centering
    \makebox[\linewidth][c]{\includegraphics[width=1\linewidth]{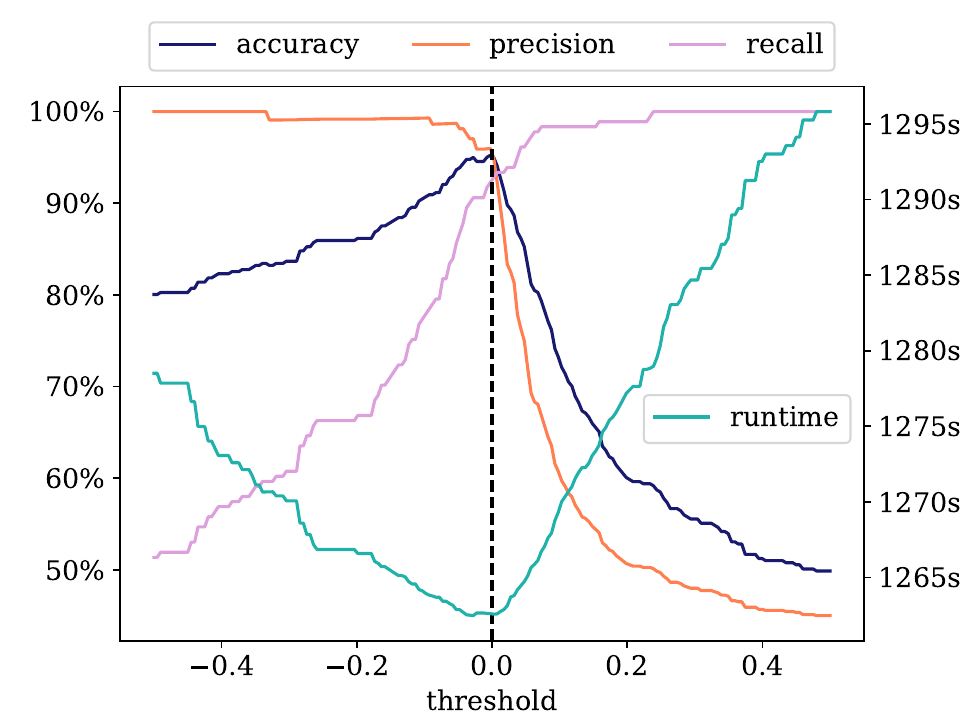}}
    \caption{Accuracy, precision, recall, as well as the e2e runtime over the test set, of the regression model converted to a classifier, depending on the threshold set (decision tree regression, PostgreSQL, all queries).}
    \label{fig:regr_threshold}
\end{figure}




\subsection{Fine-tuning precision and recall}

\change{So far we have only considered choosing a {\em threshold} of the predicted runtime difference of the regression model at $0$ (i.e., if the 
predicted runtime difference is below this threshold, we choose  Yannakakis-style query evaluation,  
and, otherwise, we
opt for the original query execution).
However, we can make use of the regression model as a useful tool to configure the trade-off between precision and recall, depending on the requirements of the application. To simplify the presentation, we focus on one DBMS and on one model, namely PostgreSQL and decision trees. 
The extension to other models and DBMSs is straightforward and yields similar results. 
\rp{Es waere vermutlich schon sinnvoll, die entsprechenden Figures fuer die anderen DBMSs im Appendix zu bringen.}
In Figure \ref{fig:regr_threshold}, we show how changing the threshold affects the accuracy, precision and recall 
of the resulting classification model. Clearly, the maximum accuracy is at the $0$-threshold. The accuracy, however, continues to be high in the direction of negative thresholds, while falling off quickly in the direction of a positive threshold. 
In particular, this shows that an optimization of precision can be performed without sacrificing much recall. }

\change{ To summarize our findings on the quality of predicting when Yannakakis-style query evaluation is better: the results from \Cref{tab:ml.performance_class} and \Cref{tab:ml.performance_reg} that show very high levels of accuracy, precision and recall for our chosen model of decision trees, as well as the ability to further optimize for precision as seen in \Cref{fig:regr_threshold} allow us to positively answer the key question \textbf{Q1}.
In \Cref{fig:regr_threshold} we also get an answer for our key question \textbf{Q4}: by choosing the right threshold one can achieve almost perfect precision, with only modest reductions in accuracy and e2e runtime (which we discuss in detail in \Cref{subsec:effects_perf}).
}


\subsection{Insights from Decision Trees}

\begin{table}[t]

\setlength{\tabcolsep}{3pt}
\centering
\caption{Most important features according to Gini coefficient of the Decision Tree models. The features are described in Section \ref{sect:FeatureSelection}}
\label{tab:gini}
\begin{tabular}{lc||lc}
\toprule
\multicolumn{2}{c}{PostgreSQL} & \multicolumn{2}{c}{DuckDB} \\
\textbf{Feature} & \textbf{Gini} & \textbf{Feature}  & \textbf{Gini}\\
\midrule
\textbf{P3}$^*$ max(est. join rows) & 0.369 & \textbf{B1}\phantom{$^*$} is 0MA? & 0.280 \\
\textbf{B1}\phantom{$^*$} is 0MA? & 0.157 & \textbf{B7}$^*$ mean(cont. c.) & 0.206 \\
\textbf{P2}$^*$ q75(est. sing. table rows)	 & 0.069 & \textbf{D1}$^*$ max(est. card.) & 0.147  \\
\textbf{P1}\phantom{$^*$} est. total cost & 0.053 & \textbf{D1}$^*$ q25(est. card.) & 0.061 \\
 \textbf{P3}$^*$ min(est. join rows) & 0.051 & \textbf{D1}$^*$ q75(est. card.) & 0.057\\
\bottomrule
\end{tabular}
\end{table}

\begin{figure*}[h!]
\definecolor{Gray}{RGB}{155,155,155}
\definecolor{LightGray}{RGB}{180,180,180}
\definecolor{BaseA}{RGB}{173,216,230}        
\definecolor{BaseB}{RGB}{135,206,235}        
\definecolor{RewritingA}{RGB}{144,238,144}   
\definecolor{RewritingB}{RGB}{60,179,113}    
\definecolor{AIA}{RGB}{255,182,193}          
\definecolor{AIB}{RGB}{255,160,122}          

\makeatletter
\newcommand\resetstackedplots{
\makeatletter
\pgfplots@stacked@isfirstplottrue
\makeatother
\addplot [forget plot,draw=none] coordinates{(Base,0) (Rewriting,0) (SMASH,0)};
}
\begin{tikzpicture}
  \begin{groupplot}[
    group style={
      group size=3 by 1,         
      horizontal sep=1.5cm,        
    },
    ybar stacked,
    /pgf/bar width=28pt,
    ymax=6000,
    ymin=0,
    height=5.5cm,
    width=6cm,
    symbolic x coords={Base, Rewriting, SMASH},
    xtick={},
    nodes near coords align={vertical},
    enlarge x limits=0.3,      
    ymajorgrids=true,
    grid style=dashed,          
    legend style={
      at={(7.5,-0.5)},            
      anchor=north,
      /tikz/every even column/.append style={column sep=0.5cm}
    },        
    legend to name=named,
    legend columns=6,
    legend cell align={left}
    ]
\nextgroupplot[
    ylabel={e2e-time (s)},
      title={\large PostgreSQL}
    ]
    
        
      \addplot+ [thick, fill=BaseA, draw=black] coordinates {(Base,3124)};
      \addplot+ [thick, fill=BaseB, draw=black] coordinates {(Base,525)};
      
      \resetstackedplots
      \addplot+ [ thick, fill=RewritingA, draw=black] coordinates {(Rewriting,923)};
      \addplot+ [ thick, fill=RewritingB, draw=black] coordinates {(Rewriting,737)};
      \resetstackedplots

      \addplot+ [ thick, fill=AIA, draw=black] coordinates {(SMASH,760)};
      \addplot+ [ thick, fill=AIB, draw=black] coordinates {(SMASH,503)};

   \node at (axis cs:Base,3965) [anchor=south, font=\small] {3965};
    \node at (axis cs:Rewriting,1849) [anchor=south, font=\small] {1849};
    \node at (axis cs:SMASH,1256) [anchor=south, font=\small] {1256};

    \nextgroupplot[
      title={\large SparkSQL}
    ]
        
      \addplot+ [forget plot,thick, fill=BaseA,draw=black] coordinates {(Base,5246)};
      \addplot+ [forget plot,thick, fill=BaseB,draw=black] coordinates {(Base,595)};
      \resetstackedplots
      \addplot+ [forget plot,thick, fill=RewritingA,draw=black] coordinates {(Rewriting,1110)};
      \addplot+ [forget plot,thick, fill=RewritingB,draw=black] coordinates {(Rewriting,1052)};
      \resetstackedplots
      \addplot+ [forget plot,thick, fill=AIA,draw=black] coordinates {(SMASH,1047)};
      \addplot+ [forget plot,thick, fill=AIB,draw=black] coordinates {(SMASH,580)};

 \node at (axis cs:Base,5241) [anchor=south, font=\small] {5841};
    \node at (axis cs:Rewriting,2162) [anchor=south, font=\small] {2162};
    \node at (axis cs:SMASH,1627) [anchor=south, font=\small] {1627};
    
    \nextgroupplot[
      title={\large DuckDB}
    ]
        
      \addplot+ [thick, fill=BaseA,draw=black] coordinates {(Base,2580)};
      \addplot+ [thick, fill=BaseB,draw=black] coordinates {(Base,81)};
      \resetstackedplots
      \addplot+ [thick, fill=RewritingA,draw=black] coordinates {(Rewriting,1190)};
      \addplot+ [thick, fill=RewritingB,draw=black] coordinates {(Rewriting,634)};
      \resetstackedplots
      \addplot+ [thick, fill=AIA,draw=black] coordinates {(SMASH,531)};
      \addplot+ [thick, fill=AIB,draw=black] coordinates {(SMASH,71)};

     \node at (axis cs:Base,2661) [anchor=south, font=\small] {2661};
    \node at (axis cs:Rewriting,1824) [anchor=south, font=\small] {1824};
    \node at (axis cs:SMASH,602) [anchor=south, font=\small] {602};

    \legend{Base 0MA, Base Enum, Rewriting 0MA, Rewriting Enum, SMASH 0MA,SMASH Enum}
    
  \end{groupplot}

  \ref{named}

\end{tikzpicture}

\caption{Comparison of e2e performance over the test set queries for three database systems: PostgreSQL, Spark (via SparkSQL), and DuckDB. The full bar indicates the runtime over all test set queries, the lower mark indicates the time for 0MA queries. 
}
\label{fig:main.dbperf}
\end{figure*}
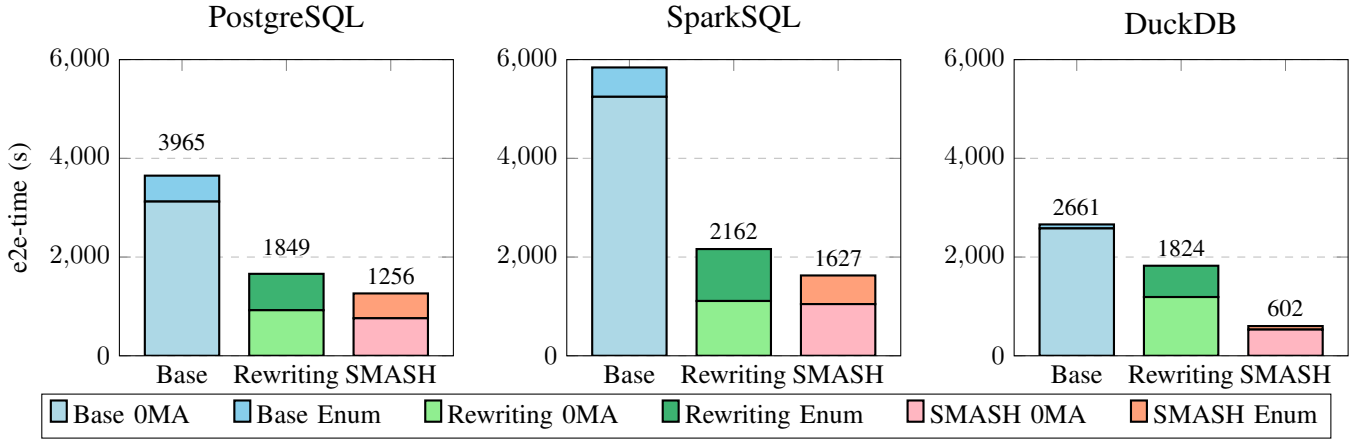

It is of particular interest that decision trees are among the top-performing models as they are highly interpretable and can provide us with deeper insights into the features that strongly affect the prediction.  In Table \ref{tab:gini}, we present the Gini coefficients of the top 5 most relevant attributes, for the PostgreSQL and the DuckDB decision tree classifiers
\ifArxiv
(for details on SparkSQL and further attributes, see Table \ref{tab:giniext} in the Appendix).
\fi
\change{Note that \emph{the feature set between the two systems is different}, as explained in \Cref{sect:FeatureSelection}, so we cannot compare the Gini coefficients of all the same features.}
The Gini coefficient~\cite{breiman2017classification}
measures the contribution of the feature to the outputs of the model. Looking at the Gini coefficients, we see that, although the performance of the models trained on the PostgreSQL and DuckDB features and runtimes are very similar, the decision trees rely on different features. In the case of DuckDB, the feature indicating whether the query is a 0MA query has the highest importance. While the PostgreSQL model strongly relies on the maximum join rows in the plan, DuckDB's second-most-important feature is the mean number of container counts, a feature extracted from the join tree. 
\change{The model for PostgreSQL, on the other hand, only use a single basic feature, namely \textbf{B1} {(is 0MA?)}, among the most important features.}
\change{
This ability of decision trees to highlight which specific features affect the prediction the most, helps us answer the key question \textbf{Q2} positively. We hope our full experimental artifacts, which include the full decision trees, will help foster further research into using these features to improve query engine optimization in general. }





\subsection{Effects on Database Performance}
\label{subsec:effects_perf}

So far, we have evaluated purely the performance of the machine learning models on the dataset that we created for our algorithm selection problem. However, our initial hypothesis was that  machine learning based algorithm selection can solve the complex challenge of deciding when evaluation in the style of Yannakakis' algorithm is preferable. We therefore now move on to evaluating the performance of the resulting full system that uses our trained algorithm selection models to decide when to rewrite, and thus execute queries using the predicted best query execution method.

\change{
We first need to decide which of the two models, the decision tree model or the regression model, we pick for the experiments, since we aim to identify the best algorithm selection mechanism. However, looking at \Cref{tab:ml.performance_reg}, which shows the performance of the regression model, and \Cref{tab:ml.performance_class}, which shows the performance of the classification model, we see that the two are fairly comparable, with a slight advantage for the regression model. As we also saw, the regression-based approach is especially promising for real-world applications as it allows us to fine-tune the decision threshold. We thus choose the regression approach with a threshold at $0$ for the analysis presented in this section.

}


As mentioned in the introduction, we will refer to this integrated method, that decides whether to rewrite to Yannakakis-style evaluation based on the prediction of the decision tree model as 
\ourTool, short for \emph{\textbf{S}upervised \textbf{M}achine-learning for \textbf{A}lgorithm \textbf{S}election  \textbf{H}euristics}.
In this section, we will perform all experiments only on the test set queries. \emph{No queries from these experiments were seen by the model at training time, inluding to select the best model.}

At the most fundamental level, we are interested in improving overall query answering performance. We investigate this by analyzing the end-to-end (e2e) 
time necessary to answer all queries in the test set, \change{where ``end-to-end'' refers to taking the time of running all the benchmark queries and looking at it cumulatively. We note that this also includes the time for the algorithm selection included in SMASH, which was around 2 milliseconds end-to-end. }


We summarize our analysis in \Cref{fig:main.dbperf}, where \textit{Base} refers to the baseline of executing the queries directly in the DBMS, \textit{Rewriting} refers to the time always using the rewriting to Yannakakis-style evaluation, and \ourTool refers to the use of our algorithm selection model as described above. To study the robustness of our approach we perform these experiments on three different DBMSs: PostgreSQL, Spark (via SparkSQL), and DuckDB. The significant technical differences between the three systems provide us with a way to study the performance of our method independently of specific DBMS technologies.
\change{  
We report timeouts as follows: if only one of the evaluation methods (rewriting or base case) times out, 
then we report in \Cref{fig:main.dbperf} for such queries as runtime the value of the timeout (= 100s).
%
On the other hand, if both evaluation methods time out, we exclude them from the comparison, since algorithm selection cannot affect anything in such a case. 
Out of the 441 queries involved in the test set, there were 27 queries that timed out for both evaluation methods on  PostgreSQL, 
13 queries that timed out for both evaluation methods on  SparkSQL, and 30 queries that timed out for both evaluation methods on DuckDB.  
}

Consistently over all systems, we can observe a large improvement over both alternatives by using algorithm selection. Furthermore, we see that even always rewriting overall causes significant improvements over baseline execution of all three systems tested. However, this improvement comes from speedups specifically on queries that are hard for traditional RDBMS execution. These large improvements offset more common minor slowdowns using the \textit{Rewriting} approach. Using \ourTool we are able to get the best of both worlds, the major speedups without the minor slower cases. 

\change{We conclude that, for our key question \textbf{Q3} regarding the effect of the quality of algorithm selection on query evaluation 
times,  we can give a positive answer. 
Indeed, our algorithm selection clearly improves the e2e query evaluation times across a number of different queries and three different DBMSs. }

\begin{figure}[t]
\begin{tikzpicture}
\begin{axis}[
    /pgfplots/bar cycle list/.style={/pgfplots/cycle list={%
                {brown!60!black,fill=brown!30!white,mark=none,thick},%
                {black,fill=gray,mark=none, thick},%
                },
    },
    ybar,
    bar width=15pt,
    enlarge x limits=0.3, 
    ymax = 70,
enlarge y limits = 0.05,
    legend style={
        at={(0.5,1.04)},
        anchor=south,
        legend columns=-1
    },
    width=8.5cm,
    height=4.5cm,
    ylabel={Slowdown cases (\%)},
    symbolic x coords={PostgreSQL,SparkSQL,DuckDB},
    xtick=data,
    nodes near coords={\pgfmathprintnumber\pgfplotspointmeta\%},
    nodes near coords align={vertical},
    ymajorgrids=true,
]

\addplot coordinates {(PostgreSQL,59) (SparkSQL, 63) (DuckDB,63)}; 

\addplot coordinates {(PostgreSQL,2) (SparkSQL, 2) (DuckDB,3)}; 

\legend{Rewriting, SMASH}

\end{axis}
\end{tikzpicture}
\caption{Analysis of how often a query is slower than the base case when always rewriting to Yannakakis-style execution vs. when rewriting depending on our trained algorithm selection models (out of 441 queries in the test set).}
\label{fig:slowdown}
\end{figure}
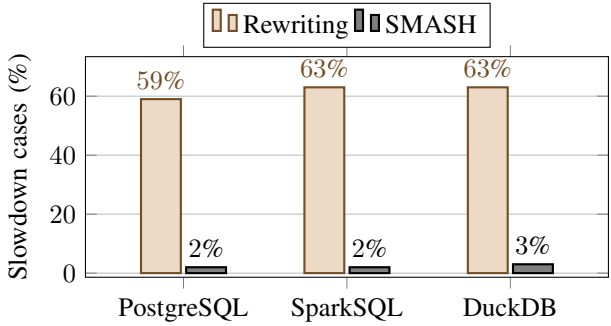

To provide further insight into how e2e query evaluation performance is affected, \Cref{fig:slowdown} shows the percentage of queries in which \textit{Rewriting} and \textit{\ourTool} are slower than \textit{Base}. While the rewriting approach yields large improvements in e2e performance, it also causes minor slowdown of queries in over half of all cases. The data illustrates clearly that \ourTool provides a convincing solution to the problem, with minor slowdowns on only around 2\% of 
the queries, combined with even larger improvements in e2e performance.

\subsection{The effect of Data Augmentation.}

We explore the impact of data augmentation on model performance through a focused ablation study, carefully examining the effectiveness of our augmentation strategies. Specifically, we compare models trained on two distinct training sets: one encompassing the fully augmented dataset derived from the complete \ourBm{}, and another restricted solely to \textit{base} queries with all augmented data removed. Table \ref{tab:aug_effect} succinctly summarizes this comparison, presenting accuracy, precision, and recall metrics for both the "Base" set comprised of 0MA and enumeration queries without filter augmentation, and the full augmented training set. Our analysis reveals that incorporating data augmentation substantially enhances accuracy and recall, with an even more pronounced improvement observed in precision.

\begin{table}[t]
    \centering
    \caption{Ablation study comparing the performance of a model on a training set with augmented data to one without. Values are based on the evaluation of the regression model with a 0.5-threshold on the test set.}
    \label{tab:aug_effect}
    \begin{tabular}{l||ccc|ccc}
        \toprule
        & \multicolumn{3}{c}{Base} & \multicolumn{3}{c}{Augmented} \\
        \midrule
        \textbf{DBMS} & Acc. & Prec. & Rec.
        & Acc. & Prec. & Rec.
        \\
        \midrule
       Postgres & 0.88 & 0.83 & 0.89 
       & 0.95 & 0.95 & 0.93 
       \\
       DuckDB & 0.86 & 0.79 & 0.84 
       & 0.91 & 0.88 & 0.87 
       \\
       SparkSQL & 0.85 & 0.79 & 0.81 
       & 0.93 & 0.92 & 0.88 
       \\
        \bottomrule
    \end{tabular}
\end{table}

\subsection{Significance of Performance Improvements}
The results presented in this section, particularly the runtimes shown in Figure~\ref{fig:main.dbperf}, demonstrate the clear effectiveness of our decision procedure in selecting the best evaluation method. For completeness, we also conducted statistical significance tests to rigorously confirm that these improvements stem from the proposed optimization techniques and are not the result of random variation.

We compared mean and median runtimes of our decision procedure against always using the original DBMS method and always applying Yannakakis-style evaluation. Since the comparisons are performed on the same test set, we employed statistical tests that account for such dependencies. For median comparisons, we used the Wilcoxon signed-rank test~\cite{Wilcoxon1945}, and for mean comparisons, the paired sample t-test~\cite{Ross2017}.
\ifArxiv
Further details of these tests are provided in \Cref{subsec:further_details_significance}.
\fi

Here, we simply report that all computed p-values are well below the common threshold of 0.005, clearly establishing that our results are statistically significant. In answering our key question \textbf{Q3}, we can thus clearly observe that the performance improvements achieved by our algorithm selection method are indeed \textit{significant}.

\section{Conclusion}
\label{sect:conclusion}

In this paper, we tackled a persistent and fundamental challenge in query optimization: many techniques that promise significant improvements deliver real-world performance gains only in \textit{some}, but not \textit{all}, scenarios. Our study focused on Yannakakis-style query evaluation, a method with strong theoretical foundations that aims to reduce unnecessary intermediate results through semi-joins. Nevertheless, despite this promising foundation, Yannakakis-style evaluation is often slower than conventional two-way join trees in practice, depending on the specific query and data characteristics.

To unlock this potential, we framed the choice between Yannakakis-style optimization and conventional DBMS evaluation as an algorithm selection problem. Our decision procedure, grounded in empirical evidence across multiple popular RDBMSs, delivers substantial performance improvements and demonstrates that learning when to apply an optimization is just as critical as the optimization itself.

Although we focused on acyclic and 0MA queries, the methodology we propose is broadly applicable. It opens the door to more intelligent optimization strategies across a wide range of techniques, including decomposition-based methods and cyclic queries. In particular, we plan to combine our approach with recent advances in hypertree decomposition-based optimizations~\cite{lanzinger2024soft}, which have shown promising but variable performance across different workloads. In such settings, a principled algorithm selection framework is key to maximizing their practical impact.

\clearpage
\ifArxiv
\textbf{Acknowledgements. }
This work has been funded by the Vienna Science and Technology Fund (WWTF) [10.47379/ICT2201,10.47379/VRG18013].
\fi






\bibliographystyle{ieeetr}
\bibliography{main}


\ifArxiv
\clearpage
\input{edbt-appendix}
\fi

\end{document}

%% file: edbt-appendix.tex

\appendix


\section{Further Information for Section~\ref{sect:DataAugmentation}}
\label{app:dataAugmentation}

In this section we aim to give further details into the data augmentation. In \Cref{ex:data_augment1}, we have
illustrated the filter augmentation. We continue here by adding another example to illustrate enumeration augmentation.


\begin{example} \label{ex:data_augment3}

Consider the query $q_{\mathit{aug}1}$, given in \Cref{ex:data_augment1}. We present three different enumeration augmentations, whereby the query output is transformed from producing a single output to \emph{enumerating} a number of rows. 
One option to ``enum-augment'' $q_{\mathit{aug}1}$ is to swap  the aggregation function \mintinline[escapeinside=||,fontsize=\small]{sql}{MIN(v.Id)}, with a projection on the {\tt Id} column of the {\tt users} table, and the {\tt UserID} column of the {\tt votes} table producing the query $q^{\mathit{e}1}_{\mathit{aug}1}$. Other options, shown below, lead to the queries $q^{\mathit{e}2}_{\mathit{aug}1}$ and $q^{\mathit{e}3}_{\mathit{aug}1}$.

{
\begin{minted}[escapeinside=||,fontsize=\small,fontfamily=tt]{sql}

|{\color{gray}$q_{\mathit{aug}1}$:}|  SELECT MIN(u.Id) 
       FROM   votes AS v, badges AS b, users AS u 
       WHERE  u.Id = v.UserId AND v.UserId = b.UserId 
              AND v.BountyAmount>=40 AND v.BountyAmount<=50 
              AND u.DownVotes=0
              
|{\color{gray}$q^{\mathit{e}1}_{\mathit{aug}1}$:}|  SELECT |\tt{\textbf{u.Id, v.UserId}}| 
         FROM  votes AS v, badges AS b, users AS u 
         WHERE u.Id = v.UserId AND v.UserId = b.UserId 
               AND v.BountyAmount>=40 AND v.BountyAmount<=50 
               AND u.DownVotes=0
               
|{\color{gray}$q^{\mathit{e}2}_{\mathit{aug}1}$:}| SELECT |\tt{\textbf{u.Id, b.UserId}}| 
         FROM  votes AS v, badges AS b, users AS u 
         WHERE u.Id = v.UserId AND v.UserId = b.UserId 
               AND v.BountyAmount>=40 AND v.BountyAmount<=50 
               AND u.DownVotes=0
               
|{\color{gray}$q^{\mathit{e}3}_{\mathit{aug}1}$:}| SELECT |\tt{\textbf{v.UserId, b.UserId}}|
         FROM  votes AS v, badges AS b, users AS u 
         WHERE u.Id = v.UserId AND v.UserId = b.UserId 
               AND v.BountyAmount>=40 AND v.BountyAmount<=50 
               AND u.DownVotes=0
\end{minted}
}

\end{example}

\section{Details on the Query Rewriting}
\label{app:query_rewriting_example}

\begin{figure}[t]
  \centering 
    \includegraphics[width=.45\textwidth]{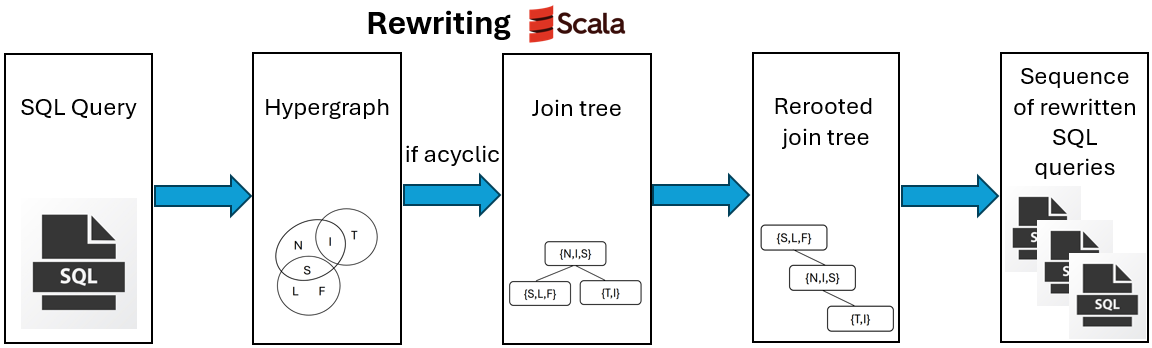} 
  \caption{Workflow for rewriting the queries.}  
  \label{fig:rewriting}  
\end{figure} 

In this section we aim to give the reader technical insights into the query rewriting process, where we follow the approach of~\cite{DBLP:journals/corr/abs-2303-02723}.
This approach involves rewriting each query into an equivalent sequence of SQL queries (i.e. one that gives the same result) but which ``forces'' (or, at least, guides)  
the DBMSs to apply a Yannakakis-style evaluation strategy.
The rewrite process is visualized in Figure~\ref{fig:rewriting}.

We thus take a SQL query as input and first transform it into a hypergraph. 
The approach by \cite{DBLP:journals/corr/abs-2303-02723} begins by applying the GYO-reduction~\cite{DBLP:conf/compsac/YuO79}.  One thus verifies that the CQ is acyclic and, if so, constructs a join tree. \change{This approach does not support cyclic queries, since they do not permit join trees.}
In case of 0MA queries, containing an aggregate expression with a single attribute, one takes the relation of this single  attribute as the root of the join tree, as this allows to significantly simply the Yannakis' algorithm by skipping multiple traversals of the join tree, thus allowing for a more efficient evaluation. The rewriting proceeds by creating a sequence of SQL queries. One traverses the tree in a bottom-up fashion, starting from the leafs, and produces auxillary tables via {\tt{\color{OliveGreen}\textbf{CREATE VIEW}}}, for the leaf nodes, and {\tt{\color{OliveGreen}\textbf{CREATE UNLOGGED TABLE}}}, for the inner nodes of the join tree.
For 0MA queries the aggregation is performed last, after the creation of auxiliary tables.

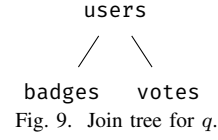
\begin{figure}
    \centering
    \scalebox{1}{
    \begin{forest}
    for tree={align=center}
        [{\texttt{users} 
        }            
            [{\texttt{badges} 
            }]
            [{\texttt{votes} 
            }]
        ]
    \end{forest}
    }
    \vspace{-5mm}
    \caption{Join tree for $q$.}
    \label{fig:tpch-query-jointree-app}
    \vspace{-5mm}
\end{figure}

We illustrate the query rewrite process by revisiting the 0MA query $q$ from \Cref{ex:data_augment1}.
To make it easier to follow, we provide the join tree of $q$ in \Cref{fig:tpch-query-jointree-app} and also repeat the definition of $q$ itself.

\begin{minted}[escapeinside=||,fontsize=\small,fontfamily=tt]{sql}

|\color{gray}$q$|: SELECT MIN(u.Id) 
        FROM votes as v, badges as b, users as u 
        WHERE u.Id = v.UserId AND v.UserId = b.UserId 
             AND v.BountyAmount>=0 AND v.BountyAmount<=50 
             AND u.DownVotes=0

|\color{gray}$q_{\mathit{r}1}$|: CREATE VIEW E3 AS SELECT * 
        FROM users AS users
        WHERE users.DownVotes = 0
        
|\color{gray}$q_{\mathit{r}2}$|: CREATE VIEW E2 AS SELECT *
        FROM badges AS badges
        
|\color{gray}$q_{\mathit{r}3}$|: CREATE |\tt{\color{OliveGreen}\textbf{UNLOGGED}}| TABLE E3E2 AS SELECT *
        FROM E3 WHERE EXISTS (SELECT 1
                    FROM E2 
                    WHERE E3.Id=E2.UserId)
                    
|\color{gray}$q_{\mathit{r}4}$|: CREATE VIEW E1 AS SELECT * 
        FROM votes AS votes
        WHERE CAST(votes.BountyAmount AS INTEGER) >= 0 
             AND CAST(votes.BountyAmount AS INTEGER) <= 50
             
|\color{gray}$q_{\mathit{r}5}$|: CREATE |\tt{\color{OliveGreen}\textbf{UNLOGGED}}| TABLE E3E2E1 AS
        SELECT MIN(Id) AS EXPR$0
        FROM E3E2 WHERE EXISTS (SELECT 1
                    FROM E1
                    WHERE E3E2.Id=E1.UserId)
                    
|\color{gray}$q_{\mathit{r}6}$|: SELECT * FROM E3E2E1
\end{minted}


In the rewriting of the example query, we can see that users and badges are joined first, even if they did not have an equality condition together in the original query. This is due to the fact that the three relations are joined on the same attribute and, after 
the replacement of equi-joins by natural joins, the distinction between explicit and implicit equality condition is irrelevant (and also not visible anymore). The use of {\tt{\color{OliveGreen}\textbf{UNLOGGED TABLE}}} is an optimization to guide the DBMS to avoid writing the intermediate views to disk.

We note that the rewriting also produces a sequence of DROP statements to delete the created tables after the evaluation.




\section{Further Information for Section \ref{sect:FeatureSelection}}
\label{app:FeatureSelection}

In this section, we aim to illustrate the complete feature vector for two example queries from our data set, namely ``{\tt HETIO\_2-01-CbGaD}'', termed $q_1$ and ``{\tt HETIO\_3-06-CdGuCtD}'' termed $q_2$, shown below.


\begin{minted}[escapeinside=||,fontsize=\small,fontfamily=tt]{sql}

|{\color{gray}$q_1$:}| SELECT MIN(c.nid) 
    FROM   compound c, binds b, gene g, associates a, disease d 
    WHERE  c.nid = b.sid AND b.tid = g.nid AND
           g.nid = a.tid AND a.sid = d.nid

|{\color{gray}$q_2$:}| SELECT MIN(c1.nid) 
    FROM   compound c1, downregulates d1, gene g,  
           upregulates u2, compound c2, treats t, disease d    
    WHERE  c1.nid = d1.sid AND d1.tid = g.nid AND
           g.nid = u2.tid AND u2.sid = c2.nid AND
           c2.nid = t.sid AND t.tid = d.nid
\end{minted}

\begin{figure}[t]
  \centering 
    \includegraphics[width=0.4\textwidth]{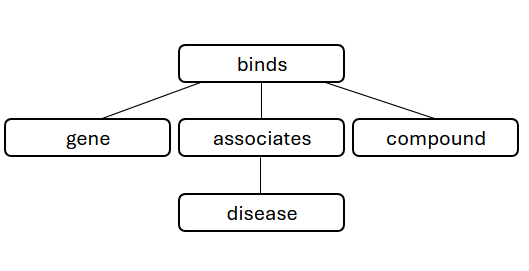} 
  \caption{Join tree of $q_1$.}  
  \label{ex_jointree1}  
\end{figure}

\begin{figure}[t]
  \centering 
    \includegraphics[width=0.45\textwidth]{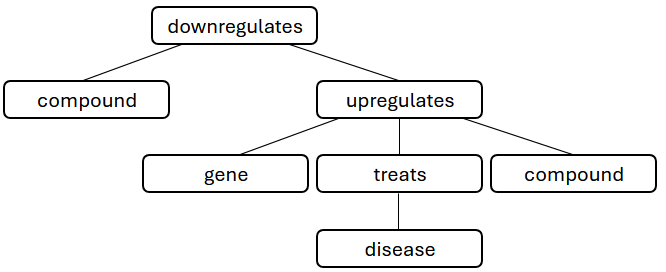} 
  \caption{Join tree of $q_2$.}  
  \label{ex_jointree2}  
\end{figure}

Their calculated join trees are given in Figures~\ref{ex_jointree1} and~\ref{ex_jointree2}, respectively. 
Recall from \Cref{sect:FeatureSelection}
that some of the features of the queries are determined by the join tree.
In Table~\ref{features}, the values of all features of these two queries are given.
For the set-based features, we also explicitly show the six statistical data points that are extracted from the set to produce the feature vector used for the models to be trained.

\section{Further Information for Section \ref{sect:SolvingAlgorithmSelectionProblem}}
\label{app:SolvingAlgorithmSelectionProblem}


We provide here a list of all hyperparameter values for the various model types considered here, i.e., 
three types of neural network models, k-NN, random forests, and SVM, and decision trees. The details are given in \Cref{tab:hyperparameters}. The experiments presented here are a continuation of the work in \cite{bohm2024rewrite}, to which we can refer for even more detailed explanations of the models and hyperparameters, as well as further literature on the machine learning concepts. In cases where mutiple hyperparameters are mentioned (such as 'kernel=linear/poly/rbf'), we have experimented with all of these. We provide a short descrition of the most important hyperparameters:

\rp{Ich wuerde den Verweis auf die Diplomarbeit (die fuer Aussenstehende auch gar nicht so leicht zu finden ist) lieber vermeiden.}
\co{Die Master thesis hat eine DOI, die ist auch im Zitat. Das es nicht leicht zu finden ist sehe ich nicht ein. Eine URL klicken zu müssen ist ja zumutbar. }
\co{Aber etwas seltsam find ich die Bemerkung schon (und nicht allein wegen dem Zitat) "is based on experiments from [10]": warum ist das nicht auf unseren Experimenten basiert? Vielleicht besser zu schreiben, dass unsere Experimente diese weiterführen, und man daher auf [10] verweisen ``darf''}
\todo[color=cyan, inline]{Alex: Ich habe den Satz jetzt, wie von Cem vorgeschlagen, etwas umformuliert}

\begin{description}
    \item[k-NN] k is the number of neighbors used by the model
    \item [Random forest] The number of estimators refers to the number of individual decision trees making up the forest
    \item [SVM] The kernel of the SVM is the function which defines the decision boundary.
    \item [MLP, HGNN, MLP+HGNN] For classification, we used cross-entropy loss, and for regression MSE (mean-squared-error). The loss function is used in the training of the model (gradient descent) to compare the actual output to the desired output. The \textit{batch size} is the number of samples used in a single training step.
    The number of \textit{epochs} refers to the number of steps used in training. \textit{Learning rate} is a parameter controlling how "fast" gradient descent adapts the weights of the model.

    The layer-configurations of the neural network models, such as 'in-40-10-out', describe how the nodes are arranged in the layers of the models. In this example, 'in' is the input layer, of in most cases 30 features, and '40' and '10' refer to the sizes of the intermediate layers. Layers are always fully connected, with ReLU activation functions.
    
\end{description}

\clearpage

\begin{table*}[t] 
\caption{A  complete overview of the features for the queries $q_1$ and $q_2$. We refer to \Cref{sect:FeatureSelection} for an explanation of each feature. Recall that each feature that consists of a set of numbers is indicated by the use of an asterix in its name e.g. B7$^*$. For these set-based features we also state the statistical data that is used in the training to produce a fixed-width vector. } 
\label{features} 
\centering 
    { 
\begin{tabular}{lcc} 
     \toprule 
      \bf Feature & $\mathbf{q_1}$ & $\mathbf{q_2}$ \\ 
      \midrule 
      {\bf B1}:\phantom{$^*$} is 0MA? & 1 & 1 \\ 
      {\bf B2}:\phantom{$^*$} number of relations & 5 & 7 \\ 
      {\bf B3}:\phantom{$^*$} number of conditions & 4 & 6 \\ 
      {\bf B4}:\phantom{$^*$} number of filters & 0 & 0  \\ 
      {\bf B5}:\phantom{$^*$} number of joins & 4 & 6 \\ 
      {\bf B6}:\phantom{$^*$} depth & 2 & 3 \\ 
      {\bf B7}$^*$: {container counts} & $\{$ \textit{1, 1, 1, 1, 1, 2, 3} $\}$ & $\{$ \textit{1, 1, 1, 1, 1, 1, 1, 1, 1, 2, 3} $\}$\\
      &   \begin{tabular}{|r|c|}
      \hline
        min &  1 \\
        max & 3  \\
        mean & 1.43 \\ 
        q25 & 1 \\ 
        med.& 1 \\ 
        q75 & 1.5 \\
        \hline
        \end{tabular} & 
  \begin{tabular}{|c|c|} 
        \hline
        min &  1 \\
        max & 3  \\
        mean & 1.27 \\ 
        q25 & 1 \\ 
        med.& 1 \\ 
        q75 & 1 \\
        \hline
        \end{tabular} \\
      
      {\bf B8}$^*$: {branching degrees} & $\{$ \textit{3, 1} $\}$ & $\{$ \textit{2, 3, 1} $\}$ \\ 

   &   \begin{tabular}{|r|c|}
      \hline
        min &  1 \\
        max & 3  \\
        mean & 2 \\ 
        q25 & 1.5 \\ 
        med.& 2 \\ 
        q75 & 2.5 \\
        \hline
        \end{tabular} &  \begin{tabular}{|c|c|} 
        \hline
        min &  1 \\
        max & 3  \\
        mean & 2 \\ 
        q25 & 1.5 \\ 
        med.& 2 \\ 
        q75 & 2.5 \\
        \hline
        \end{tabular}  \\
      
     {\bf P1}:\phantom{$^*$} estimated total cost & 1175.3 & 10283.6 \\
      
      {\bf P2}$^*$: {estimated single table rows} & $\{$ \textit{23142, 25246, 137,1, 1552} $\}$ & $\{$ \textit{154076, 1552, 146276, 1510, 137, 1552, 20945} $\}$    \\ 

 &   \begin{tabular}{|r|c|}
      \hline
        min &  1 \\
        max & 25246  \\
        mean & 10015.6 \\ 
        q25 & 137 \\ 
        med.& 1552 \\ 
        q75 & 23142 \\
        \hline
        \end{tabular} & \begin{tabular}{|c|c|} 
        \hline
        min &  137 \\
        max & 154076  \\
        mean & 46578.29 \\ 
        q25 & 1531 \\ 
        med.& 1552 \\ 
        q75 & 83610.5 \\
        \hline
        \end{tabular}  \\


      {\bf P3}$^*$: {estimated join rows} & 
      $\{$ {\textit{1190, 2361, 626, 626}} $\}$& 
      $\{$ \textit{493338, 22993, 2578, 2578, 446, 446}  $\}$ \\

 &  \begin{tabular}{|r|c|}
      \hline
        min &  626 \\
        max & 2361  \\
        mean &  1200.75\\ 
        q25 & 626 \\ 
        med.& 908 \\ 
        q75 & 1482.75 \\
        \hline
        \end{tabular} &  \begin{tabular}{|c|c|} 
        \hline
        min &  446 \\
        max & 493338  \\
        mean & 87063.17 \\ 
        q25 & 979 \\ 
        med.& 2578 \\ 
        q75 & 17889.25  \\
        \hline
        \end{tabular} \\
      

      {\bf D1}$^*$: {estimated cardinalities} & 
      $\{$ \textit{31829, 29051, 26334, 26334,}   &
     $\{$ \textit{4993749, 354275, 338789, 322069,2299, }\\

      & \textit{418, 2299, 29051, 24035,26334} $\}$ & 
      \textit{2299, 338789, 2090, 24035,2299, 322069, 418, 2090} $\}$ \\
 &  \begin{tabular}{|r|c|}
      \hline
        min &  418 \\
        max & 31829  \\
        mean &  21742.78 \\ 
        q25 & 24035 \\ 
        med.& 26334 \\ 
        q75 & 29051 \\
        \hline
        \end{tabular} &  \begin{tabular}{|c|c|} 
        \hline
        min &  418 \\
        max & 4993749  \\
        mean & 515790 \\ 
        q25 & 2299 \\ 
        med.& 24035 \\ 
        q75 & 338789  \\
        \hline
        \end{tabular} \\
      
      \bottomrule 
    \end{tabular} 
}
  \vspace{3cm}
\end{table*}

\clearpage

\begin{table*}[t]
\caption{ An overview of all hyperparameters of the models used during training. For the MLP model, the meaning of ``in'' in the layer column refers to the number of features, and the world ``out'' refers to the number of classes (2,3 and 1 for regression). 
} 
\label{tab:hyperparameters} 
\centering 

\begin{tabular}{ccc} 
     \toprule
	\textbf{model} & \textbf{hyperparameter} & \textbf{layer} \\ 
    \midrule
	k-NN & k=5 & -  \\ 
    \midrule
	Decision tree & - & - \\ 
    \midrule
	Random forest & n\_estimators=100 & - \\
    
    \midrule
	SVM & kernel=linear/poly/rbf & -\\ 
    
    \midrule
	\multirow{20}{*}{MLP} & \multirow{20}{*}{\parbox{4.5cm}{\centering Loss=Cross-Entropy/MSE \\ Batch size=100 \\ Epochs=300 (saving best model) \\ Learning rate=0.1}} & in-5-out\\ 
	 & & in-10-out\\ 
	 & & in-20-out\\ 
	 & & in-25-out\\ 
	 & & in-40-out\\ 
	 & & in-60-out\\ 
	 & & in-10-5-out\\ 	
	 & & in-20-10-out\\ 
	 & & in-40-20-out\\ 
	 & & in-40-10-out\\ 
	 & & in-60-40-out\\ 
	 & & in-60-20-out\\ 
	 & & in-80-50-out\\ 
	 & & small median, best MLP\\ 
	 & & small mean, best MLP\\ 
	 & & small min, best MLP\\ 
	 & & small max, best MLP\\ 
	 & & small q25, best MLP\\ 
	 & & small q75, best MLP\\ 
	 & & custom, best MLP\\ 
     
    \midrule
	\multirow{5}{*}{HGNN} & \multirow{5}{*}{\parbox{4.5cm}{\centering Loss=Cross-Entropy/MSE \\ Epochs=100 (saving best model) \\ Learning rate=0.001 \\ Max-Pooling}} & kernel 3x3, 1-16-32-out\\ 
	 & & kernel 3x3, 1-32-16-out\\ 
	 & & kernel 3x3, 1-16-32-16-out\\ 
	 & & kernel 3x3, 1-32-64-out\\ 
	 & & kernel 3x3, 1-4-16-out\\ 
    \midrule
	\multirow{5}{*}{combined} & \multirow{5}{*}{\parbox{4.5cm}{\centering Loss=Cross-Entropy/MSE \\ Epochs=100 (saving best model) \\ Learning rate=0.001 \\ Max-Pooling}} & best MLP-2/best HGNN-2/4-out \\  
	 &  & best MLP-5/best HGNN-5/10-out \\  
	 &  & best MLP-5/best HGNN-5/10-20-out \\  
	 &  & best MLP-10/best HGNN-10/20-40-2 \\  
	 &  & best MLP-10/best HGNN-10/20-60-20-2 \\ 

    \bottomrule
    \end{tabular}

\end{table*}

\clearpage

\section{Further Details on the Experimental Evaluation}
\label{app:AdditionalExp}


We present here further information on the experiments conducted in \Cref{sect:evaluationAlgorithmSelection}. 

\paragraph{Full details on Gini coefficients.} Recall that in \Cref{sect:evaluationAlgorithmSelection} we mention the role of Gini coefficients and list only the five features with the highest coefficients for PostgreSQL and DuckDB.  As was mentioned, the Gini coefficient, as used in this paper, measures the contribution of the feature to the outputs of the model.
In \Cref{tab:giniext} we then give a complete overview of the 8 highest Gini coefficients among the features of the trained decision tree models for all three DBMSs, including SparkSQL.


\begin{table}[t]
\setlength{\tabcolsep}{3pt}

\small
\centering
\caption{Most important features according to Gini coefficients of the Decision Tree models.}
\label{tab:giniext}
\begin{tabular}{l||lc}
\toprule
\parbox[t]{2mm}{\multirow{9}{*}{\rotatebox[origin=c]{90}{PostgreSQL}}}  &
\textbf{Feature} & \textbf{Gini} \\
\midrule
&max(est join rows) & 0.369 \\
&is 0MA? & 0.157  \\
&q75(est. sing. table rows)	 & 0.069 \\
&total cost & 0.053 \\
&min(est. sing. join rows) & 0.051  \\
&q25(est. sing. join rows)	& 0.044 \\
&median(est. join rows)	& 0.033  \\
&q25(est. single. table rows) & 0.033 \\
\end{tabular}
\begin{tabular}{l||lc}
\toprule
\parbox[t]{2mm}{\multirow{9}{*}{\rotatebox[origin=c]{90}{DuckDB}}}  &
\textbf{Feature} & \textbf{Gini}\\
\midrule
&is 0MA? & 0.280  \\
& mean(cont. c.) & 0.206  \\
& max(est. card.) & 0.147  \\
&q25(est. card.) & 0.061  \\
&q75(est. card.) & 0.057  \\
&median(est. card.) & 0.047  \\
& mean(est. card.) & 0.040 \\
& min(est. card.) & 0.037 \\
\end{tabular}

\begin{tabular}{l||lc}
\toprule
\parbox[t]{2mm}{\multirow{9}{*}{\rotatebox[origin=c]{90}{SparkSQL}}}  &
\textbf{Feature}  & \textbf{Gini}\\
\midrule
&is 0MA? & 0.246 \\
&max(est. sing. join rows) & 0.212 \\
&\#joins & 0.135 \\
& est. total cost & 0.092 \\
&q25(est. sing. table rows)  & 0.053 \\
& median(est. sing. table rows) & 0.044 \\
&max(est. sing. table rows)	 & 0.037 \\
& mean(cont. c.) & 0.032 \\
\bottomrule
\end{tabular}

\end{table}

\paragraph{Further Details on the Significance Tests.}
\label{subsec:further_details_significance}
We briefly commented in \Cref{sect:evaluationAlgorithmSelection} on the significance tests we performed. Here we give a more detailed report, including explanations on how these tests are defined.


For our significance tests, we compare the mean or median of the runtimes achieved
when applying our decision procedure versus always using the original evaluation method of the DBMSs and
always using Yannakakis-style evaluation, respectively. 

Since we compare runtimes obtained with the same test set, we need to use statistical tests, which take these dependencies into consideration. For the median we take the Wilcoxon sign-rank test~\cite{Wilcoxon1945} and for the mean we use a paired sample t-test~\cite{Ross2017}.

\smallskip
\noindent
\textit{Wilcoxon sign-rank test.} The null hypothesis of this test for two (dependent) groups A and B is that the medians are equal: $H_0: median(A) = median(B)$. To get the test statistic the differences between all pairs of group A and B are calculated and ranked. Additionally, the sign of the difference is used, so that all ranks of the positive differences are summed and the same for the negative ones. The minimum of these two is the test statistic, which then can be compared to the Wilcoxon signed rank table to get the p-value. If the p-value is smaller than a chosen alpha-level, the null can be rejected and the two cases lead to significantly different medians.

\medskip
\noindent
\textit{Paired sample t-test.} The null hypothesis of this test for two (dependent) groups $A$ and $B$ is that the means are equal: $H_0: mean(A) = mean(B)$. Again, the differences of the pairs of values are used to calculate the test statistic. In this case it is a t-test statistic with $n-1$ degrees of freedom and looks like the following.
\begin{align*}
t=\frac{\bar{d} \sqrt{n}}{s}, \qquad \text{with } \bar{d} = \frac{1}{n} \sum\limits_{i=1}^n d_i, s=\sqrt{\frac{1}{n-1} \sum\limits_{i=1}^n (d_i-\bar{d})^2} 
\end{align*}
Here the t-test tables can be used to get the p-value and again if it is smaller than alpha, the null can be rejected and we can conclude that the means are significantly different.

For both tests, we choose 0.1 as alpha value as a common choice. 

In Table~\ref{tab:signifianceVsOriginalEvaluation}, we have the p-values for the 
Wilcoxon sign-rank test (Wilcoxon s.-r.) in the left group of 2 columns with numbers and for the 
paired sample t-test (p.s. t-test) in the 
right-most  group of 2 columns. Inside each group, the first column compares the 
median resp. mean of the runtimes obtained by the original evaluation method (Orig) of each DBMS with 
the evaluation method chosen by the decision procedure (Dec). The second column of each group
compares the
median resp. mean of the runtimes obtained by the rewritten queries (Rewr, i.e., by Yannakakis-style query  evaluation) with 
the evaluation method chosen by the decision procedure (Dec). 
We can see that all p-values are extremely small. For any reasonable choice of alpha value (note that
alpha values such as 0.1 or 0.05 are common choices; the p-values in the table are much smaller) we can reject the null hypothesis in all cases. This means that the decision procedure indeed yields a 
significant performance improvement (in terms of means and median of runtimes). 

\begin{table}[t]

\setlength{\tabcolsep}{3pt}
\caption{p-values resulting from the Wilcoxon sign-rank test and paired sample t-test, comparing the 
original evaluation method of each DBMS with the decision program.}
\label{tab:signifianceVsOriginalEvaluation}
\centering

\begin{tabular}{c|c|c}
\toprule
& \multicolumn{2}{c|}{\textbf{Median (Wilcoxon s.-r.)}}\\ \midrule
\textbf{System} & \textbf{Orig/SMASH} & \textbf{Rewr/SMASH}  \\ \midrule
PostgreSQL & $1.1\cdot10^{-28}$ & $1.5\cdot10^{-41}$  \\
Spark & $1.8\cdot10^{-23}$ & $4.8\cdot10^{-48}$  \\
DuckDB & $7.2\cdot10^{-22}$ & $2.1\cdot10^{-42}$ \\


\end{tabular}

\begin{tabular}{c|c|c}
\toprule
& \multicolumn{2}{c}{\textbf{Mean (p.s. t-test)}} \\ \midrule
\textbf{System} & \textbf{Orig/SMASH} & \textbf{Rewr/SMASH}  \\ \midrule
PostgreSQL &  $1.0\cdot10^{-10}$ & $0.0002$ \\
Spark &  $6.9\cdot10^{-12}$ & $1.8\cdot10^{-7}$ \\
DuckDB &  $2.0\cdot10^{-8}$ & $2.2\cdot10^{-5}$\\

\bottomrule

\end{tabular}

\end{table}

\begin{table}[t]

\setlength{\tabcolsep}{2pt}
    \caption{Runtime breakdown by benchmark for PostgreSQL.}
    \label{tab:e2e-breakdown}
    \centering
    \begin{tabular}{l|llllll}
    \toprule
    \bf   & \bf \bf \multirow{3}{*}{\rotatebox[origin=t]{-45}{Accur.}} & \bf e2e &  \bf e2e & \bf e2e & \bf e2e & \bf  \\ 
    \bf Benchmark &   & \bf orig & \bf rewr & \bf SMASH & \bf opt & \bf \phantom{a} \# \\ 
     & & (sec) & (sec) & (sec) &   \\
    \midrule
         HETIO & \phantom{0}94.5\%     & \phantom{0}701   & 198 &  192  & 192 & \phantom{0}73\\
         STATS & \phantom{0}94.8\%     & \phantom{0}938   & 335 & 308  & 307 & 306 \\
         SNAP & 100.0\%       & 1978  & 733 & 731  & 731 & \phantom{0}21 \\
         JOB & \phantom{0}97.5\%       & \phantom{00}23    & 384 & \phantom{0}23   & \phantom{0}23 & \phantom{0}40 \\
         LSQB & 100.0\%       & \phantom{000}9     & \phantom{00}9 & \phantom{00}9    & \phantom{00}9 & \phantom{00}1 \\
        \bottomrule
    \end{tabular}

\end{table}